\documentclass[a4paper]{article}
\pdfoutput=1 

\usepackage{setspace}
\usepackage{jheppub}
\usepackage[T1]{fontenc}
\usepackage[latin9]{inputenc}
\usepackage{graphicx}

\newcommand{\noun}[1]{\textsc{#1}}
\usepackage[mathscr]{euscript}

\def\KZ#1 {{\bf  \textcolor{blue}{[KZ: {#1}]}}}

\def\pt{\mathrm{p_{\scriptscriptstyle{T}}}}
\def\ptww{\mathrm{p_{\scriptscriptstyle{T,W^+W^-}}}}
\def\ptmiss{\mathrm{p_{\scriptscriptstyle{T,miss}}}}
\def\ptmin{\mathrm{p_{\scriptscriptstyle{T,min}}}}
\def\ptj{\mathrm{p_{\scriptscriptstyle{T,j_1}}}}
\def\ptjj{\mathrm{p_{\scriptscriptstyle{T,j_2}}}}

\newcommand{\POWHEG}{\noun{Powheg}}
\newcommand{\NLOPS}{\noun{Nlops}}
\newcommand{\PYTHIAEIGHT}{\noun{Pythia 8}}
\newcommand{\POWHEGBOX}{\noun{Powheg-Box}}
\newcommand{\MINLO}{\noun{Minlo}}
\newcommand{\MINLOp}{\noun{Minlo}$^\prime$ }
\newcommand{\WWJMINLO}{\noun{WWj-Minlo}}
\newcommand{\WW}{\noun{WW}}
\newcommand{\ww}{$W^{\scriptscriptstyle{+}}W^{\scriptscriptstyle{-}}$ }
\newcommand{\wwplusj}{$W^{\scriptscriptstyle{+}}W^{\scriptscriptstyle{-}}\!+\rm{jet}$ }

\preprint{\\\\CERN-TH-2016-146\\LAPTH-031/16\\OUTP-16-17P\\}

\title{Merging $WW$ and $WW$+jet with M\Large{INLO}}

\author[a]{Keith Hamilton,}
\author[b,c]{Tom Melia,}
\author[d]{Pier Francesco Monni,}
\author[e]{Emanuele Re,}
\author[d,f]{Giulia Zanderighi}

\affiliation[a]{Department of Physics and Astronomy, University College London,
\\London, WC1E 6BT, UK}

\affiliation[b]{Berkeley Center for Theoretical Physics, University of California, Berkeley, CA 94720}

\affiliation[c]{Theoretical Physics Group, Lawrence Berkeley National Laboratory, Berkeley, CA 94720}

\affiliation[d]{Rudolf Peierls Centre for Theoretical Physics, University of Oxford, Oxford OX1 3NP, UK}

\affiliation[e]{LAPTh, Universit\'e Savoie Mont Blanc, CNRS, B.P.110,
  Annecy-le-Vieux F-74941, France}

\affiliation[f]{Theoretical Physics Department, CERN, Geneva, Switzerland}

\emailAdd{keith.hamilton@ucl.ac.uk}
\emailAdd{tmelia@lbl.gov}
\emailAdd{pier.monni@physics.ox.ac.uk}
\emailAdd{emanuele.re@lapth.cnrs.fr}
\emailAdd{giulia.zanderighi@cern.ch}

\abstract{We present a simulation program for the production of a pair
  of $W$ bosons in association with a jet, that can be used in
  conjunction with general-purpose shower Monte Carlo generators,
  according to the \POWHEG{} method. We have further adapted and
  implemented the \MINLO$^\prime$ method on top of the NLO calculation
  underlying our \wwplusj generator. Thus, the resulting simulation
  achieves NLO accuracy not only for inclusive distributions in
  \wwplusj production but also \ww production, i.e. when the
  associated jet is not resolved, without the introduction of any
  unphysical merging scale. This work represents the first extension
  of the \MINLO$^\prime$ method, in its original form, to the case of
  a genuine underlying $2\to2$ process, with non-trivial virtual
  corrections.  }


\begin{document}
\maketitle
\flushbottom

\section{Introduction}

Run II at the LHC will further explore physics at the scale of
electroweak symmetry breaking and continue the search for new
phenomena in the TeV energy range. Within this programme, major
attention will be paid to sharpening measurements of Higgs boson
properties, searching for direct and indirect signals of new
particles, and excluding and curtailing the form of proposed physics
beyond the standard model.
For all of these endeavours, the QCD production of pairs of
electroweak (EW) gauge bosons is an important process over which to
exert theoretical and experimental control. Weak boson pair production
constitutes a major background in Higgs boson analyses, as well as in
direct searches for new particles decaying into jets, leptons and/or
missing energy. Furthermore, precision measurements of these processes
translate to indirect bounds on new physics at higher energies than
are directly accessible, e.g. through setting constraints on the
allowed size of anomalous trilinear gauge interactions.

Studies of \ww hadroproduction have been carried out both at the
Tevatron\footnote{See
  e.g.~\cite{Aaltonen:2009aa,CDFnote,Abazov:2012ze} and references
  therein.} and the LHC,\footnote{See
  e.g.~\cite{Aad:2016wpd,Khachatryan:2015sga} and references therein.}
in which limits on anomalous triple gauge couplings were derived. For
LHC Run I measurements, experimental uncertainties are at the level of
7-8\%, and are dominated by systematics.
One important source of uncertainty occurs when both $W$ bosons decay
leptonically: the final-state contains two neutrinos, whose missing
momenta prevent a full reconstruction of the event kinematics---in
particular the momenta of the $W$ bosons.  The lack of any resonant
$W$ mass peaks leads to a greater sensitivity of experimental analyses
to the theoretical modelling of this process than would otherwise be
the case, be it as a signal, or a background.  The kinematic
distributions which are used as experimental handles have a greater
susceptibility to QCD radiative corrections; the uncertainty connected
to this modelling is a contributing factor in the experimental
systematic error estimate.
Separately, in order to isolate \ww final-states from backgrounds, in
particular those due to $t\bar t$ and $tW$ production, experimental
analyses categorise events according to their jet multiplicity, in
so-called \emph{jet-binned}/\emph{jet-vetoed} analyses.  Such event
selections are employed in the study of both QCD continuum \ww pair
production, as well as Higgs boson production in the $H\rightarrow
W^{\scriptscriptstyle{+}}W^{\scriptscriptstyle{-}}$ channel (in which
context the former signal process plays the r\^{o}le of an irreducible
background). In both analyses, the fact that the signal definition
includes cuts on the presence of associated jets also implies a
potentially marked sensitivity to higher order QCD effects: in the
study of continuum QCD \ww production the ensuing uncertainty
significantly contributes to the extrapolation to the total
cross-section.
All considered, the above experimental issues point to the importance
of flexible, high-accuracy, and fully realistic, theoretical
predictions for \ww and jet-associated \ww production processes.

The level of theoretical precision with which hadronic $W$-pair
production is known has seen truly remarkable progress in recent
years. Partonic QCD calculations for $pp\to
W^{\scriptscriptstyle{+}}W^{\scriptscriptstyle{-}}$ have evolved from
stable-$W$ approximations at LO \cite{Brown:1978mq} to much more
sophisticated treatments, incorporating spin correlation and off-shell
effects in $W$ decays, all at
NLO~\cite{Dixon:1998py,Campbell:1999ah,Dixon:1999di}. The latter are
available as flexible public computer codes, such as
MCFM~\cite{Campbell:2011bn}.
The NLO calculation of $W$-pair production in association with a jet,
including $W$ decays, was first carried out almost ten years
ago~\cite{Campbell:2007ev}, with the dijet case following in
2011~\cite{Melia:2011dw}. Gluon initiated contributions to $W$-pair
hadroproduction were calculated in refs.~\cite{Dicus:1987dj,
  Glover:1988fe, Binoth:2005ua, Binoth:2006mf}, the contribution due
to interference with Higgs boson production being later taken into
account in ref.~\cite{Campbell:2011cu}.  The leading order gluon
fusion contribution to jet-associated $W$-pair production was first
computed in ref.~\cite{Melia:2012zg}.
More recently, in the last couple of years, NNLO predictions for
$W$-pair production have become available, for the case of stable
bosons~\cite{Gehrmann:2014fva}, using the two loop helicity amplitudes
of ref.~\cite{Gehrmann:2014bfa}. Off-shell two-loop amplitudes have
also been computed lately, both for quark-antiquark
collisions~\cite{Caola:2014iua,Gehrmann:2015ora}, and gluon
fusion~\cite{Caola:2015ila,vonManteuffel:2015msa}. The latter results
have been used in determining the NLO corrections to $gg\to
W^{\scriptscriptstyle{+}}W^{\scriptscriptstyle{-}}$~\cite{Caola:2015rqy}.
The quark-antiquark two-loop amplitudes have recently been used to
present fully differential NNLO predictions \cite{Grazzini:2016ctr}.
Finally, at fixed order, we note that NLO electroweak corrections to
$W$-pair production are also
known~\cite{Bierweiler:2012kw,Baglio:2013toa,Billoni:2013aba}, even
including a full off-shell treatment of the $W$
decays~\cite{Biedermann:2016guo}.

In the context of resummed QCD calculations, transverse momentum and
threshold resummations for this process have been studied in
refs.~\cite{Grazzini:2005vw,Wang:2013qua,Meade:2014fca,Grazzini:2015wpa}
and ref.~\cite{Dawson:2013lya} respectively. The effects of a jet veto
resummation were considered by several groups
\cite{Jaiswal:2014yba,Becher:2014aya,Monni:2014zra,Dawson:2016ysj}, in
part triggered by a discrepancy between the measured and predicted
cross sections at the LHC.

Monte Carlo event generators matching NLO $W$-pair production
calculations to parton showers (\noun{Nlops}) have been publicly
available for around a decade. Indeed this process was the subject of
the pioneering proof-of-concept work demonstrating the MC@NLO
\noun{Nlops} matching formalism ~\cite{Frixione:2002ik}. Alternative
\noun{Nlops} implementations of this process, using different methods
and approximations, were subsequently implemented in the
\noun{Herwig++}~\cite{Hamilton:2010mb},
\noun{Sherpa}~\cite{Hoche:2010pf} and
\noun{Powheg-Box}~\cite{Melia:2011tj,Nason:2013ydw} packages. The
original \noun{Herwig++} implementation has recently been superseded
by a version including single-resonant and gluon induced
contributions, in the new \noun{Herwig7}
framework~\cite{Bellm:2016cks,Bellm:2015jjp}.  \noun{Nlops}
simulations of $W$-pair production and $W$-pair production in
association with a jet have been merged by the \noun{OpenLoops+Sherpa}
group~\cite{Cascioli:2013gfa} according to the
\noun{Meps@Nlo}~\cite{Gehrmann:2012yg,Hoeche:2012yf} merging scheme.
The \noun{aMC@NLO} team have also presented state-of-the-art
simulations of NLO weak boson pair production in
ref.~\cite{Frederix:2011ss}, which can be automatically merged with
higher order jet multiplicities according to the \noun{FxFx} multi-jet
merging method~\cite{Frederix:2012ps,Alwall:2014hca}. All of the above
simulations include full leptonic decay kinematics, with the latter
two simulations (employing NLO merging) also incorporating $gg$
initiated contributions.
 
In this paper we extend and apply the \MINLOp technique of
ref.~\cite{Hamilton:2012rf}, to deliver a NLO calculation of
jet-associated $W$-pair production, which is simultaneously NLO
accurate in the description of 0-jet quantities.
In our underlying NLO calculation the $W$ bosons are allowed to decay
either hadronically or leptonically. In the hadronic case, we
incorporate the NLO QCD correction to the decay only inclusively.
We employ fully off-shell matrix elements, including singly-resonant
contributions, but we omit the loop-mediated $gg$ channel in our
simulation. The latter is separately finite and can be accounted for
straightforwardly via, e.g., the \noun{gg2ww} event
generator~\cite{Binoth:2006mf,Kauer:2012hd} (as employed by ATLAS and
CMS).
The enhanced \MINLOp computation is implemented within the
\noun{Powheg-Box} \noun{Nlops}
framework~\cite{Nason:2004rx,Frixione:2007vw,Alioli:2010xd}, such that
approximate higher order, higher twist, and non-perturbative QCD
effects can be accounted for (parton shower, hadronization and
underlying event), rendering a realistic description of the
final-state. The latter class of corrections can have a non-negligible
impact on many observables; in particular, for jet-binned cross
sections, where they contribute to migrations between bins, their
effects can be sizable, as noted in, e.g., ref.~\cite{Astill:2016hpa}.

We reiterate that not only is our single \MINLOp calculation of
jet-associated $W$-pair production capable of populating the phase
space of the 0-jet region, it can be formally proved that the
predictions which it yields for 0-jet quantities are also NLO
accurate. This is in difference to the earlier-mentioned multi-jet
merged \noun{Nlops} simulations, which essentially partition phase
space into jet bins, whose `size' is set by a new \emph{merging scale}
parameter, with each bin being populated by events from an
\noun{Nlops} simulation with the corresponding jet multiplicity.

The current work is novel theoretically, in so far as it represents
the first application of the original \MINLOp method to a genuine
$2\to2$ colour singlet process at the lowest order,\footnote{An
  alternative extension of \MINLOp was given in
  ref.~\cite{Frederix:2015fyz}.} wherein the virtual ($V$) and Born
($B$) contributions are not proportional to each other. The ratio
$V/B$ enters the process-dependent part of the NNLL resummation
coefficient $B_2$, whose inclusion in the \noun{Minlo} Sudakov form
factor is mandatory for yielding NLO accuracy also in the description
of 0-jet quantities.
In the original \MINLOp works on Higgs and Drell-Yan production the
ratio $V/B$ is just a constant (since these are effectively $2\to1$
processes) while the extension to the present case requires a
procedure to compute the process-dependent $B_2$ term for the
production of a generic colour-singlet system.

The \MINLOp implementation presented here can be readily promoted to a
\noun{Nnlops} simulation of \ww production following the same
procedure employed to build \noun{Nnlops}
$H$~\cite{Hamilton:2013fea,Hamilton:2015nsa},
$Z$~\cite{Karlberg:2014qua}, and $HW$~\cite{Astill:2016hpa}
generators.
The present work can be regarded as a main theoretical step towards
such a \noun{Nnlops} simulation of \ww production.

An \noun{Nnlops} generator could also be achieved through
matching an $\rm{NNLO+NNLL}^{\prime}$ resummed calculation of this 
process to a parton shower using the \noun{Geneva} matching 
formalism~\cite{Alioli:2012fc,Alioli:2013hqa,Alioli:2015toa,Alioli:2016wqt}.
In addition, it would appear to be a straightforward matter to merge the same
NNLO calculation with a parton shower according to the
$\rm{UN}^{2}\rm{LOPS}$ prescription~\cite{Hoeche:2014aia,Hoche:2014dla}.

The paper is structured as follows. In section \ref{sec:Method} we
give details on the construction of our underlying NLO calculation for
$W$-pair production in association with a jet, as well as details on
the validation of our implementation. We then proceed to exemplify the
extension of the \MINLOp method to a generic colour-singlet
process. Many aspects of the \MINLOp approach follow unchanged from
refs.~\cite{Hamilton:2012np, Hamilton:2012rf}, and so we focus on
presenting the key differences and their practical implementation.
Section~\ref{sec:Validation-and-results} presents a phenomenological
study of kinematic distributions for the decay more of the $W$-bosons
to $e^+\nu_e\mu^-\overline{\nu}_\mu$ case. We summarize our findings
and conclude in sect.~\ref{sec:Conclusion}.
We have made our simulation publicly available within the
\noun{Powheg-Box} code.\footnote{Instructions to download the code can
  be obtained at \url{http://powhegbox.mib.infn.it}. }

\section{Method and technical details\label{sec:Method}}

In this section we first give all details concerning the construction
of the pure \NLOPS{} simulation of jet-associated $W$-pair production
(henceforth \noun{WWj}), including the treatment of heavy fermions and
the CKM matrix. We subsequently detail the validation of this
construction.
Following this, we go on to describe how we have modified and extended
the original \MINLOp method, such that our \noun{WWj-Minlo} simulation
also recovers NLO accurate results for 0-jet and inclusive $W$-pair
production observables (henceforth \noun{WW}).
%


\subsection{\noun{Nlops} construction\label{sub:NLOPS-construction}}

We have generated Born and real matrix elements using the \POWHEGBOX{}
interface to \noun{Madgraph 4}~\cite{Alwall:2007st} developed in
ref.~\cite{Campbell:2012am}.
The virtual matrix elements have been obtained using \noun{GoSam
  2.0}~\cite{Cullen:2014yla}.
Our code is based on matrix elements for the following Born sub-processes
and all of their associated NLO
counterparts:~\footnote{In Sec.~\ref{sec:nlovalidation} we will also discuss the impact of removing
  the gauge-invariant set of fermionic loop corrections.}
\begin{equation}
q \bar q \to e^+\nu_e\mu^-\overline{\nu}_\mu g\,,\qquad \qquad
q g \to e^+\nu_e\mu^-\overline{\nu}_\mu q\,,\qquad \qquad
\bar q g \to e^+\nu_e\mu^-\overline{\nu}_\mu \bar q\,.
\label{eq:process}
\end{equation}
Hence, while we refer to our simulation as being one of \noun{WWj}
production, we do in fact include all related off-shell and
single-resonant contributions.

We have chosen to work throughout in the four-flavour scheme (4FNS),
as employed, for instance, in the NNLO calculation of \ww production
in ref.~\cite{Gehrmann:2014fva}. Thus, we do not include effects from
third generation quarks. In doing so we most easily avoid significant
complications that affect 5FNS calculations, in particular those due
to the opening of resonant $tW$ and $t\bar{t}$ channels at ${\cal
  O}(\alpha_{\scriptscriptstyle{\rm{S}}})$ and ${\cal
  O}(\alpha_{\scriptscriptstyle{\rm{S}}}^{2})$. The latter resonant
top-pair contributions enhance the inclusive $pp\to
W^{\scriptscriptstyle{+}}W^{\scriptscriptstyle{-}}+X$ cross section by
a factor of 4 (8) at the 7 (14) TeV LHC~\cite{Gehrmann:2014fva}, but
they give rise to experimentally separable signatures.  This
necessitates a theoretical definition of \ww hadroproduction wherein
the top contributions are subtracted, analogous to that employed for
experimental measurements of this process. This issue has been studied
in ref.~\cite{Gehrmann:2014fva}, where it was shown that by an
appropriate removal of the resonant top contributions in the 5FNS
case, the 4FNS (with third generation quarks omitted) and the 5FNS
NNLO predictions agree at the level of 1-2\%.

For the $W$ boson decays in our program, the user can select leptonic
decay modes (summed over generations, or for just a single
generation), and/or hadronic decay modes (summed over all
kinematically allowed flavours).
The chosen decay channels are then taken into account when generating
Les Houches events. When a leptonic decay into more than one
generation is selected, we randomly generate the lepton flavour,
accounting for the relevant combinatorial factors. For a hadronic
decay, an up or charm quark is chosen at random and the related
down-type quark is selected with a probability given by the associated
Cabibbo matrix element squared. When writing out the corresponding
part of the Les Houches event record, we assign the quark and
anti-quark originating from the decay of the $W$ to the same colour
line. Furthermore, for hadronic decays we include the NLO correction
to the inclusive $W$-hadronic branching ratio.\footnote{In the case of
  hadronic decays, we neglect t-channel boson exchanges.} In the case
of decays to leptons of the same family, we do not include double
resonant $ZZ$ production with one boson decaying to leptons and one
invisibly.  In fact, we consider the latter as being part of the $ZZ$
production process, and, as shown in ref.~\cite{Melia:2011tj}, the
interference between the $WW$ and $ZZ$ mediated processes is
completely negligible.

We defer details of the technical checks performed in the course of
assembling this NLO calculation, within the \noun{Powheg-Box}
framework, to sect.~\ref{sec:nlovalidation}, where a full validation
of our final \noun{WWj-Minlo} generator is given.

\subsection{NLO calculation validation\label{sub:NLO-calculation-validation}}
\label{sec:nlovalidation}

The fixed order calculation described in
sect.~\ref{sub:NLOPS-construction} underlying our \noun{WWj-Minlo}
generator and implemented in the \noun{Powheg-Box} framework has been
cross-checked at leading order against
\noun{Madgraph5\_aMC@NLO}~\cite{Alwall:2014hca}, and at the NLO level
against the independent \noun{WWj} code of ref.~\cite{Melia:2010bm}.
Beside point-by-point checks of the matrix elements, total cross
sections and differential distributions were compared and found to
agree very satisfactorily in all cases.

In the calculation of one-loop matrix elements we include fermion
loops.  However, adding the latter slows down the event generation
significantly.  In our main validation work of the full
\noun{WWj-Minlo} generator, described in
sect.~\ref{sec:Validation-and-results}, we examined numerous
distributions, probing a wide range of kinematic configurations, and
did not find any distribution in which there was a statistically
significant difference exceeding 1-2\% between results obtained with
and without fermionic loop diagrams.  We therefore also release a
version of the code that omits the gauge-invariant set of fermionic
loop corrections.

\subsection{\noun{Minlo$^{\prime}$} for general jet-associated colourless
particle production processes\label{sub:Minlo-prime}}

Here we describe how to generalize the original \noun{Minlo$^{\prime}$
}procedure to deal with general jet-associated colourless particle
production, with particular reference given to \noun{WWj}. We
emphasise similarities and differences relative to the original
\emph{\noun{Minlo$^{\prime}$}} codes \cite{Hamilton:2012rf} addressing
jet-associated, \emph{single}, colourless particle production.

The \noun{Minlo$^{\prime}$} recipe and proof needs as its primary
ingredient and starting point an NLO cross section, here the
\noun{WWj} calculation described above. In general, the latter
decomposes into a sum of a part which is finite as the transverse
momentum of the colourless system ($p_{{\scriptscriptstyle
    \mathrm{T}}}$) tends to zero, $d\sigma_{{\scriptscriptstyle
    \mathcal{F}}}$, plus a correspondingly singular part,
$d\sigma_{{\scriptscriptstyle \mathcal{S}}}$. The finite piece,
$d\sigma_{{\scriptscriptstyle \mathcal{F}}}$, being power suppressed,
is essentially a spectator in proofs that \MINLOp yields NLO accuracy
for the inclusive/0-jet process, here \ww production.  The singular
part of the cross section, differential in the phase-space variables
$\Phi$ fully parametrising the underlying $q\bar{q} \to
W^{\scriptscriptstyle{+}}W^{\scriptscriptstyle{-}}$ scattering
(including W decays) and the large logarithm, $L=\ln
Q^{2}/p_{{\scriptscriptstyle \mathrm{T}}}^{2}$, here with
$Q=m_{{\scriptscriptstyle WW}}$ (the invariant mass of the
$W^{\scriptscriptstyle{+}}W^{\scriptscriptstyle{-}}$ system), can be
obtained by an explicit fixed order calculation, or expanding the NNLL
resummed $p_{{\scriptscriptstyle \mathrm{T}}}$ spectrum up to and
including $\mathcal{O}\left(\bar{\alpha}_{{\scriptscriptstyle
      \mathrm{S}}}^{2}\right)$ terms
($\bar{\alpha}_{{\scriptscriptstyle \mathrm{S}}}=\alpha_s/2\pi$). This
singular part can be obtained by identifying and replacing all
instances of the process-dependent hard function in the Drell-Yan
case, with that of \ww production.\footnote{This statement generalizes
  trivially to all jet-associated colour singlet production processes.
} The resulting expression for $d\sigma_{{\scriptscriptstyle
    \mathcal{S}}}$ thus has the form
\begin{eqnarray}
\frac{d\sigma_{{\scriptscriptstyle \mathcal{S}}}}{d\Phi dL} & = & \frac{d\sigma_{0}}{d\Phi}\,\sum_{n=1}^{2}\sum_{m=0}^{2n-1}\,H_{nm}\bar{\alpha}_{{\scriptscriptstyle \mathrm{S}}}^{n}\left(\mu_{{\scriptscriptstyle R}}^{2}\right)L^{m}\,,\label{eq:minlopr-dsigmaS}
\end{eqnarray}
where the explicit $H_{nm}$ coefficients can be extracted
from, for example, the general formulae in appendix A of
ref.~\cite{Frederix:2015fyz}; they are lengthy and so we do not repeat
them here. The process-dependent hard function, $H_{1}$, in the
$\mathcal{H}_{1}$ terms of \cite{Frederix:2015fyz}, is related to
the finite part of the renormalized NLO virtual contribution to \ww
production, $\mathcal{V}$, as follows:

\begin{eqnarray}
\mathcal{V}\left(\Phi\right) & = & \frac{1}{\Gamma\left(1-\epsilon\right)}\,\left(\frac{4\pi\mu^{2}}{Q^{2}}\right)^{\epsilon}\,\bar{\alpha}_{{\scriptscriptstyle \mathrm{S}}}\,\left[-\frac{2C_{{\scriptscriptstyle F}}}{\epsilon^{2}}-\frac{3C_{{\scriptscriptstyle F}}}{\epsilon}-C_{{\scriptscriptstyle F}}\zeta_{2}+H_{1}\left(\Phi\right)\right]\mathcal{B}\left(\Phi\right)\,.\label{eq:minlopr-renormalized-virtual-xsec}
\end{eqnarray}
Here, in eq.~\eqref{eq:minlopr-renormalized-virtual-xsec},
$\mathcal{B}$ is the Born cross section, with the normalization of
refs.~\cite{Frixione:2007vw,Alioli:2010xd}, $\mu$ is the
renormalization scale and $\epsilon$ sets the dimensionality, $d$, of
spacetime in conventional dimensional regularization
($d=4-2\epsilon$).  Apart from the one-to-one replacement of these
hard function terms, the NLO expression for
$d\sigma_{{\scriptscriptstyle \mathcal{S}}}$ is identical to the
corresponding formula for the Drell-Yan $p_{{\scriptscriptstyle
    \mathrm{T}}}$ spectrum.\footnote{In a nutshell, physically, this
  owes to the fact that the underlying primary scattering processes
  are identical from the point of view of the flow of colour charge,
  and to the universal character of the infrared QCD corrections which
  dress them to yield $d\sigma_{{\scriptscriptstyle \mathcal{S}}}$.}

Having noted the nature of the differences between the singular
behaviour of the jet-associated Drell-Yan cross section and that of
\noun{WWj}, the task of extending the \MINLOp method to the latter
reduces to that of replacing the $H_{1}\left(\Phi\right)$ function
everywhere it occurs in the procedure. Modulo this isolated change in
the recipe, the \noun{Minlo$^{\prime}$} implementation and its proofs
follow in exactly the same way as before. Indeed, in the procedure
itself the $H_{1}\left(\Phi\right)$ function only occurs once among
the additional components to be layered onto the pure \noun{WWj} NLO
calculation. Specifically, it only occurs in the process-dependent
$B_{2}$ coefficient of the \noun{Minlo$^{\prime}$} Sudakov form
factor.  Thus, the extension of the \noun{Minlo$^{\prime}$} method to
arbitrary colourless particle production processes, which are at
lowest order $q\bar q$ initiated, consists of generalizing the Sudakov
form factor exponent given in the original article
\cite{Hamilton:2012rf}:

\begin{eqnarray}
\log \Delta^{2}(Q,p_{{\scriptscriptstyle \mathrm{T}}}) & = & - \sum_{i=1}^{2} \, \int_{p_{{\scriptscriptstyle \mathrm{T}}}^{2}}^{Q^{2}}\frac{d\mu^{2}}{\mu^{2}}\,\bar{\alpha}_{{\scriptscriptstyle \mathrm{S}}}^{i}\left(\mu^{2}\right)\,\left[\,A_{i}\,\log\frac{Q^{2}}{\mu^{2}}+B_{i}\,\right]\,,\label{eq:minlops-Sudakov}
\end{eqnarray}
with
\begin{eqnarray}
A_{1} & = & 2C_{{\scriptscriptstyle F}}\,,\qquad B_{1}=-3C_{{\scriptscriptstyle F}}\,,\qquad A_{2}=2C_{{\scriptscriptstyle F}}K\,,\label{eq:minlopr-A1-B2-A2}\\
\nonumber \\
B_{2} & = & -2\gamma^{\left(2\right)}+\bar{\beta}_{0}\,C_{{\scriptscriptstyle F}}\zeta_{2}+2\left(2C_{{\scriptscriptstyle F}}\right)^{2}\zeta_{3}+\bar{\beta}_{0}H_{1}\left(\Phi\right)\,,\label{eq:minlopr-B2}
\end{eqnarray}
 and
\begin{eqnarray}
K^{\phantom{1}} & = & \left(\frac{67}{18}-\frac{\pi^{2}}{6}\right)C_{{\scriptscriptstyle A}}-\frac{10}{9}n_{{\scriptscriptstyle f}}T_{{\scriptscriptstyle R}}\,,\qquad\qquad\bar{\beta}_{0}=\frac{11C_{{\scriptscriptstyle A}}-4n_{{\scriptscriptstyle f}}T_{{\scriptscriptstyle R}}}{6}\,,\label{eq:minlopr-2-loop-cusp}\\
\nonumber \\
\gamma^{\left(2\right)} & = & \left(\frac{3}{8}-\frac{\pi^{2}}{2}+6\zeta_{3}\right)C_{{\scriptscriptstyle F}}^{2}+\left(\frac{17}{24}+\frac{11}{18}\pi^{2}-3\zeta_{3}\right)C_{{\scriptscriptstyle F}}C_{{\scriptscriptstyle A}}-\left(\frac{1}{12}+\frac{\pi^{2}}{9}\right)C_{{\scriptscriptstyle F}}n_{{\scriptscriptstyle f}}\,.\label{eq:minlopr-2-loop-non-cusp}
\end{eqnarray}
While the above formulae are specific to the case of processes which
are $q\bar{q}$ initiated at the lowest order, the necessary
modifications to deal with the $gg$ case are obvious.
Hence we see that almost the only change to be implemented in the
Drell-Yan \MINLOp code components, to enable it for use with
\noun{WWj}, is the replacement
\begin{eqnarray}
B_{2} & \rightarrow & B_{2}-\bar{\beta}_{0}H_{1}^{\left({\scriptscriptstyle \mathrm{DY}}\right)}+\bar{\beta}_{0}H_{1}^{\left({\scriptscriptstyle \mathrm{WW}}\right)}\left(\Phi\right)\,.\label{eq:minlopr-B2-shift}
\end{eqnarray}

This brings us to the final subtlety. In Drell-Yan processes, which were the subject of the original \noun{Minlo$^{\prime}$
}article \cite{Hamilton:2012rf}, $H_{1}^{\left({\scriptscriptstyle \mathrm{DY}}\right)}$
has no dependence on any kinematics and is just a number:
\begin{equation}
H_{1}^{\left({\scriptscriptstyle \mathrm{DY}}\right)}=C_{{\scriptscriptstyle F}}\left[\pi^{2}-8+\zeta_{2}\right]\,.\label{eq:minlopr-DY-etc-H1-hard-fn}
\end{equation}
This particularly simple form owes to the fact that Drell-Yan is, from
the point of virtual QCD corrections, a single, colourless, particle
production process. This is in marked contrast to the general case.
Indeed, in $W^+W^-$ production the finite virtual corrections lead to
a non-trivial dependence of $H_{1}\left(\Phi\right)$ on the kinematics
of the underlying hard scattering process which produces the primary
$W^+W^-$ system.

To effect the above transformation of $B_{2}$ in
eq.~\eqref{eq:minlopr-B2-shift}, one first needs a set of WW
kinematics, $\Phi$, with which to evaluate the
$H_{1}^{\left({\scriptscriptstyle
      \mathrm{WW}}\right)}\left(\Phi\right)$ factor. This is not a
trivial matter since in \noun{Minlo$^{\prime}$ }one has only
\noun{WWj} Born and virtual configurations, and \noun{WWjj} real
configurations. For the \noun{WWj} Born and virtual contributions we
define $\Phi$, event-by-event, by a projection of the \noun{WWj} state
onto a WW one, using the FKS mapping relevant for initial-state
radiation in NLO calculations, as described in
ref.~\cite{Frixione:2007vw}.~\footnote{We first boost all momenta to
  the frame in which the $W^+W^-$ system has zero
  rapidity. Subsequently we boost all momenta to the frame in which
  the $W^+W^-$ system has zero transverse momentum, before applying a
  final boost to the frame where the $W^+W^-$ system has the same
  rapidity it started with.}
For real emission events, we first apply a projection to the
\noun{WWj} underlying Born configuration which the \noun{Powheg-Box}
framework generated the given real configuration from in the first
place (also according to an FKS mapping), before projecting a further
step back to a \noun{WW} state in precisely the same way as described
for the \noun{WWj} Born and virtual configurations. In all cases, in
the limit that the total transverse momentum of emitted radiation is
small, the effect of the projection on the WW kinematics and its decay
products smoothly vanishes. Taking the latter two features as defining
criteria for a legitimate projection procedure, any residual
ambiguities in their definition will result only in power suppressed
corrections, affecting the precise numerical determination of $B_{2}$
safely beyond the level of the $\mathrm{N^{3}LL}_{\sigma}$ terms which
must be controlled ($B_{2}$ itself already only enters at
$\mathrm{N^{3}LL}_{\sigma}$ order); indeed, for ambiguities in the
projection to invalidate the \noun{Minlo$^{\prime}$ }simulation they
would need to give rise to at least relative
$\mathcal{O}\left(1\right)$ shifts in the numerical value of $B_{2}$
as $p_{{\scriptscriptstyle \mathrm{T}}}\to 0$. With these kinematic
considerations in hand, we compute the two $H_{1}$ terms in
eq.~\eqref{eq:minlopr-B2-shift} by calling their associated
\noun{Powheg-Box} $\mathtt{setvirtual}$ and $\mathtt{compborn}$
subroutines~\cite{Alioli:2010xd}, whose return values are
$\mathtt{virtual}$ and $\mathtt{born}$ respectively, which, for
$q\bar{q}$ initiated colour singlet production processes, we have
determined obey the following relation
\cite{Frixione:2007vw,Alioli:2010xd}:
\begin{equation}
\frac{{\tt virtual}}{{\tt born}}=H_{1}\left(\Phi\right)-C_{F}\zeta_{2}-3C_{F}\log\left(\frac{\mu_{{\scriptscriptstyle R}}^{2}}{\hat{s}}\right)-C_{F}\log^{2}\left(\frac{\mu_{{\scriptscriptstyle R}}^{2}}{\hat{s}}\right)\,,\label{eq:minlopr-virtual-over-born}
\end{equation}
 $\hat{s}$ being the invariant mass of the final-state particles.
For the case of Drell-Yan processes the latter ratio is trivially
equal to $C_{F}\left[\pi^{2}-8\right]$ for $\mu_{{\scriptscriptstyle R}}^{2}=\hat{s}$.
Finally then in practice we determine $B_{2}^{\left({\scriptscriptstyle \mathrm{WW}}\right)}$
event-by-event as 
\begin{eqnarray*}
\label{eq:B2relation}
B_{2}^{\left({\scriptscriptstyle \mathrm{WW}}\right)} & = & B_{2}^{\left({\scriptscriptstyle \mathrm{DY}}\right)}-\bar{\beta}_{0}\left[C_{F}\left[\pi^{2}-8\right]-\left.\frac{{\tt virtual}}{{\tt born}}\right|_{\mu_{{\scriptscriptstyle R}}^{2}=\hat{s}}^{\left({\scriptscriptstyle \mathrm{WW}}\right)}\right]\,.
\end{eqnarray*}

\section{Phenomenological results\label{sec:Validation-and-results}}

In this section we undertake a phenomenological study which also
serves to exemplify some of the work done to validate our
\noun{WWj-Minlo} generator, and the improved description that it
yields for a variety of important kinematic distributions. To that
end, we mainly compare our \noun{WWj-Minlo} program to the existing
\noun{Powheg-Box} WW simulation code~\cite{Melia:2011tj}. The analysis
shown in sect.~\ref{sub:Comparison-of-WW-and-WWJ-Minlo-generators} is,
however, only a representative summary of a more wide-reaching
comparative study, whose findings are mentioned in the accompanying
discussion when relevant.

\subsection{Comparison of WW and \noun{WWj-Minlo}$^{\prime}$ generators\label{sub:Comparison-of-WW-and-WWJ-Minlo-generators}}

For the purposes of validation and demonstrating the improvements
yielded by our \noun{WWj-Minlo} simulation, we compare and contrast
its predictions for a number of kinematic quantities of general
interest to those of the existing WW \noun{Powheg-Box} generator
(which we ultimately aim to replace).

In the following we consider only the process $pp\to
e^+\nu_e\mu^-\overline{\nu}_\mu+X$ and 13 TeV LHC collisions. We set
the $Z$ mass to $91.188$ GeV and its width to $2.441$ GeV. The $W$
mass and width are taken to be $80.149$ GeV and $2.0476$ GeV
respectively. We derive the value of the fine-structure constant,
according to the so-called $G_\mu$-\emph{scheme}, as being
$\alpha_{em}(M_{Z}) = 1/132.507$. The parton distribution functions we
have used are the NLO, $n_f=4$, NNPDF3.0~\cite{Ball:2014uwa} set, with
the associated running coupling, since we work in the 4FNS. Unless
otherwise stated, all predictions shown have been obtained by
showering the \noun{Powheg-Box}'s hardest emission events with
\PYTHIAEIGHT{}~\cite{Sjostrand:2006za,Sjostrand:2007gs,Sjostrand:2014zea},
including hadronization but not multi-parton interaction effects.

Both \WW{} and \WWJMINLO{} predictions are obtained with the
\noun{Powheg-Box} \texttt{bornzerodamp} feature
activated~\cite{Alioli:2010xd} (see also appendix B of
ref.~\cite{Melia:2011tj}). This flag has the effect of limiting the
amount by which the integrand of the \noun{Powheg} Sudakov form factor
exponent can depart from its soft/collinear approximation. This option
avoids potentially pathological situations wherein the Born term in
the denominator of the \noun{Powheg} Sudakov exponent enters a region
of phase space in which it, itself, is vanishing faster than the real
cross section in the numerator, when this approaches its
soft/collinear factorized form. This vanishing of tree-level matrix
elements is well known to occur in the context of, e.g., charged
current weak interactions, in which scattering processes can
`switch-off' as certain kinematical configurations are approached, due
to conflicting chirality and angular momentum
constraints.\footnote{Some such configurations are suggested in the
  context of \ww production in appendix B of the \noun{Powheg} WW
  generator paper~\cite{Melia:2011tj}.} In contrast to the WW case, in
the \noun{WWj-Minlo} generator, the effects of this setting are
generally negligible in all distributions which are not sensitive to
high transverse momentum emission of the second hardest radiated
parton in the event ($\gtrsim 200 \rm{GeV}$).  Moreover, even in such
cases, where the predictions should be considered as simply LO
accurate, the effect of the switch is modest, ranging up to 25\% in
the worst case considered here
($p_{\scriptscriptstyle{\rm{T,j_{2}}}}\sim 500 \rm{GeV}$).

The central renormalization and factorization scale choice for the
\WWJMINLO{} results is dictated by the \MINLOp
formalism~\cite{Hamilton:2012rf}. The scale choice used for the strong
coupling inside the integrand of the Sudakov form factor exponent is
the conventional setting shown in eq.~\eqref{eq:minlops-Sudakov}.  All
other instances of the strong coupling are evaluated at a scale given
by the transverse momentum of the weak-boson pair.  The factorization
scale at which the PDFs are evaluated is also set to this value. The
resummation scale used in the \noun{WWj-Minlo} Sudakov was already
given in sect.~\ref{sub:Minlo-prime} as $Q=m_{{\scriptscriptstyle
    WW}}$. For further details on the technical implementation of the
\MINLOp procedure, in particular the subtleties surrounding
renormalization and factorization scale variations, we refer the
reader to refs.~\cite{Hamilton:2012rf,Frederix:2015fyz}.  In the \WW{}
generator, the central renormalization and factorization scale choice,
for the computation of the $\bar{B}$ function, is set to the invariant
mass of the \ww system $m_{{\scriptscriptstyle WW}}$. In both
generators, the scale uncertainty bands in our plots have been
obtained by varying $\mu_{{\scriptscriptstyle R}}$ and
$\mu_{{\scriptscriptstyle F}}$, independently, up and down by a factor
of two around their central values, while keeping $\frac{1}{2} \le
\mu_{{\scriptscriptstyle R}}/\mu_{{\scriptscriptstyle F}} \le 2$.

Unless stated otherwise, all figures show comparisons of the \WW{} and
\WWJMINLO{} simulations, and are arranged pairwise. In each case the
left- (right-) hand plot shows the perturbative uncertainty of the
\WW{} (\WWJMINLO{}) calculation, and in the lower panel the ratio to
the \WW{} (\WWJMINLO{}) central result.

\subsubsection{Total cross sections\label{sub:Comparison-of-total-cross sections}}

Before launching into comparisons of differential distributions it
will help us first to understand the total inclusive cross sections of
the WW and \noun{WWj-Minlo} generators relative to one another,
including their theoretical uncertainties.
For the total inclusive \ww production cross sections we find that the
predictions of \MINLOp and the conventional WW NLO computation are in
agreement to within about 4\% obtaining $1.219_{-2.1\%}^{+2.4\%}$ pb
from the latter, and $1.174_{-4.8\%}^{+7.2\%}$ pb from \WWJMINLO{}. It
is entirely natural that the two sets of results do not agree
identically, since they differ at the level of explicit NNLO-sized
terms relative to the leading order \ww production process;
\WWJMINLO{} includes all ingredients but the two-loop virtual
corrections to the full NNLO computation, for example. Thus, we regard
the fact that the \MINLOp and conventional NLO results agree at about the
4\% level as being quite satisfactory -- indeed the agreement is slightly
better than that found previously for the case of 14 TeV
LHC Drell-Yan production in ref.~\cite{Hamilton:2012rf}.

We note that the \noun{WWj-Minlo} prediction comes with a theoretical
uncertainty (due to scale variations) which is about a factor 2-3
larger than that from the \WW{} generator.~\footnote{For the total
  inclusive cross section, the predictions of the \WW{} generator are
  identical to those of conventional NLO, owing to the exact unitarity
  of \noun{Nlops} algorithms.} This feature has already been observed
in the first \MINLOp work concerning the Drell-Yan
process~\cite{Hamilton:2012rf}, and subsequently in applying the
\MINLOp method to $HW$/$HZ$ production~\cite{Luisoni:2013kna}; in both
cases a similar-sized enlargement of the \MINLOp scale uncertainty was
found with respect to the corresponding conventional total inclusive
NLO predictions.

As in ref.~\cite{Hamilton:2012rf}, we point out that, just as one can
expect the central values of the predictions to differ, on account of
intrinsic differences at the NNLO level, one should not be surprised
to find similar-sized inequalities in the associated scale variations.
Furthermore, it is well known that scale variations often don't give a
reasonable estimate of the size of missing higher-order contributions
at LO, or even NLO. For example, in \ww production at LO there is zero
renormalization scale dependence, since there is zero dependence on
the strong coupling constant, thus the true theoretical uncertainty at
that order is significantly underestimated. For what concerns our NLO
results here though, there is very clear supporting evidence from the
NNLO studies of \ww production in ref.~\cite{Grazzini:2016ctr} that
the scale uncertainties predicted by our \noun{WWj-Minlo} program are
actually more reasonable, whereas those from conventional NLO
substantially underestimate the true error.

\subsubsection{Inclusive observables \label{sub:Comparison-of-inclusive-observables}}

The preceding observations on the WW and \noun{WWj-Minlo} total
inclusive cross sections are important to bear in mind while going on
to examine kinematic distributions, since these similarities and
differences are felt directly and indirectly by many of the plotted
quantities. Indeed, for inclusive quantities, and to some extent also
more exclusive ones, differences in normalization and uncertainty
estimates are directly attributable to those found for the total
inclusive cross sections.

\begin{figure}[htbp]
  \centering{}\includegraphics[scale=0.68]{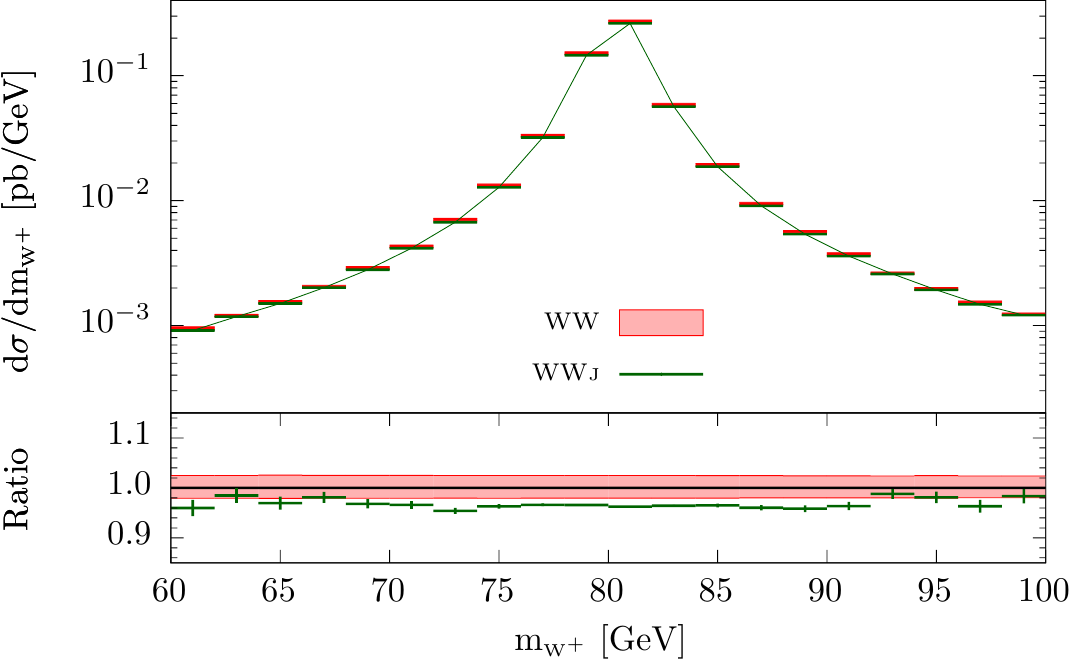}\hspace{0.1cm}\includegraphics[scale=0.68]{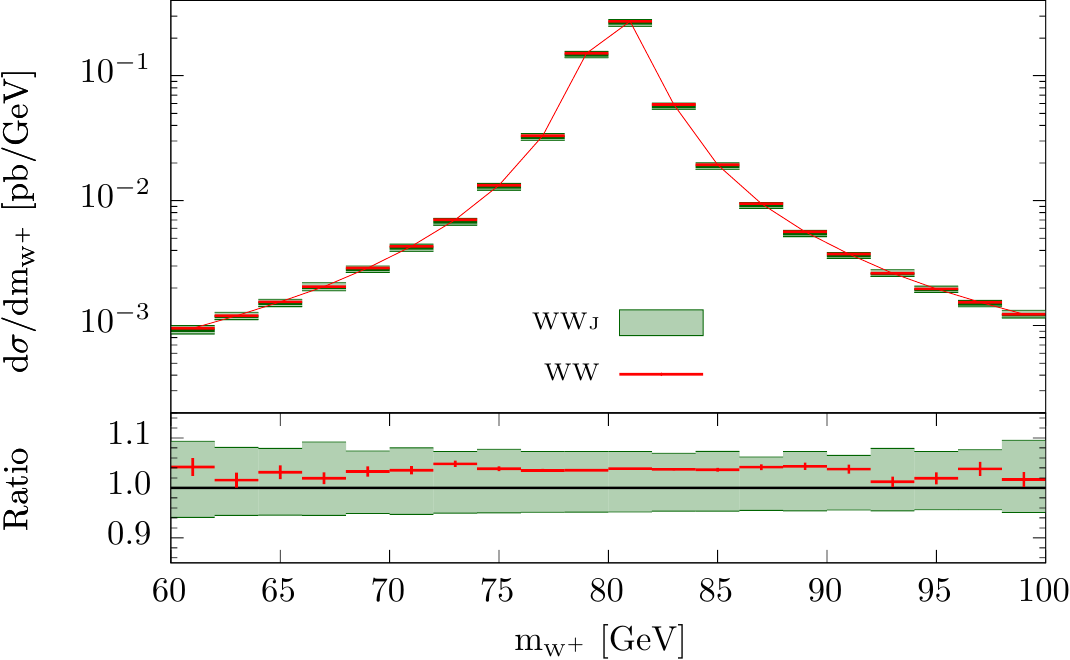}\protect\caption{\label{fig:mass-Wplus}
    The mass of the $\mathrm{W}^{+}$ boson as predicted by the
    \noun{WW} (red) and \noun{WWj-Minlo} (dark green) generators.
    All results include parton shower and hadronization corrections
    obtained by processing the \noun{Powheg-Box} hardest emission
    events with \PYTHIAEIGHT{}; MPI effects are not been included.    
}
\label{fig:mw} 
\end{figure}

We consider first differential distributions for observables which are
inclusive with respect to the presence of QCD radiation. We begin in
fig.~\ref{fig:mw} with the $W^{\scriptscriptstyle{+}}$ mass
distribution.
As expected, we see that the two generators agree rather well for this
very inclusive quantity.  The only differences between the two types
of predictions here are to do with their normalization and the width
of their scale uncertainty bands. As expected, these differences
completely reflect those seen in the corresponding total inclusive
cross sections discussed in sect.~\ref{sub:Comparison-of-total-cross
  sections}; the \WWJMINLO{} result sits at the lower edge of the
\noun{Powheg} WW uncertainty band, which lies well within the larger
\noun{WWj-Minlo} uncertainty band.

\begin{figure}[tbph]
  \centering{}\includegraphics[scale=0.68]{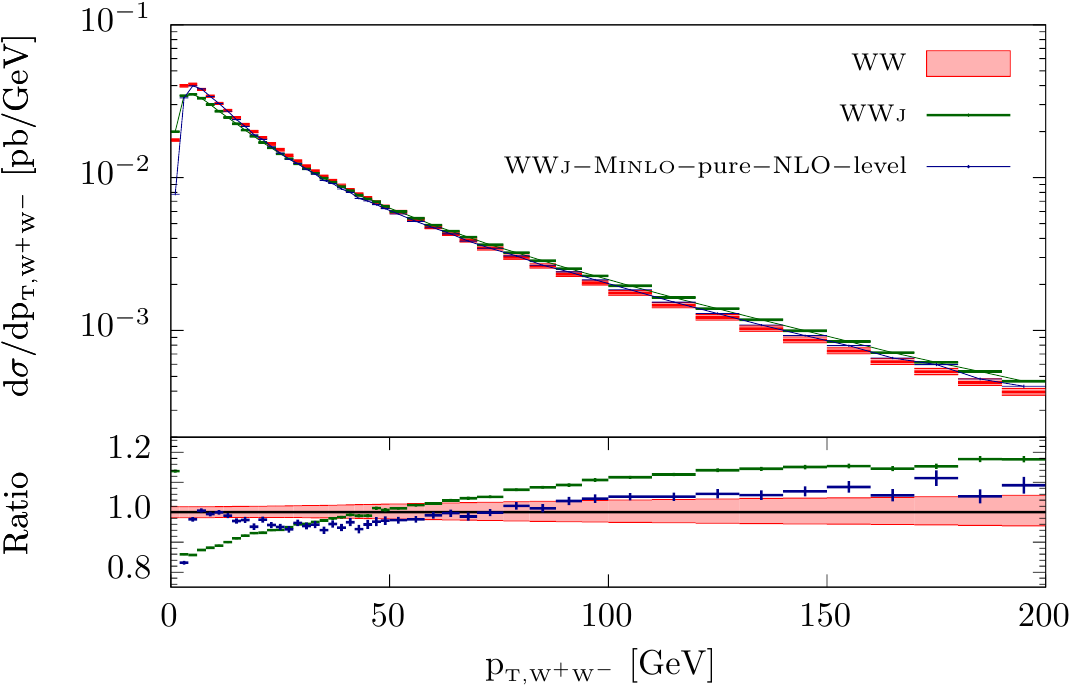}\hspace{0.1cm}\includegraphics[scale=0.68]{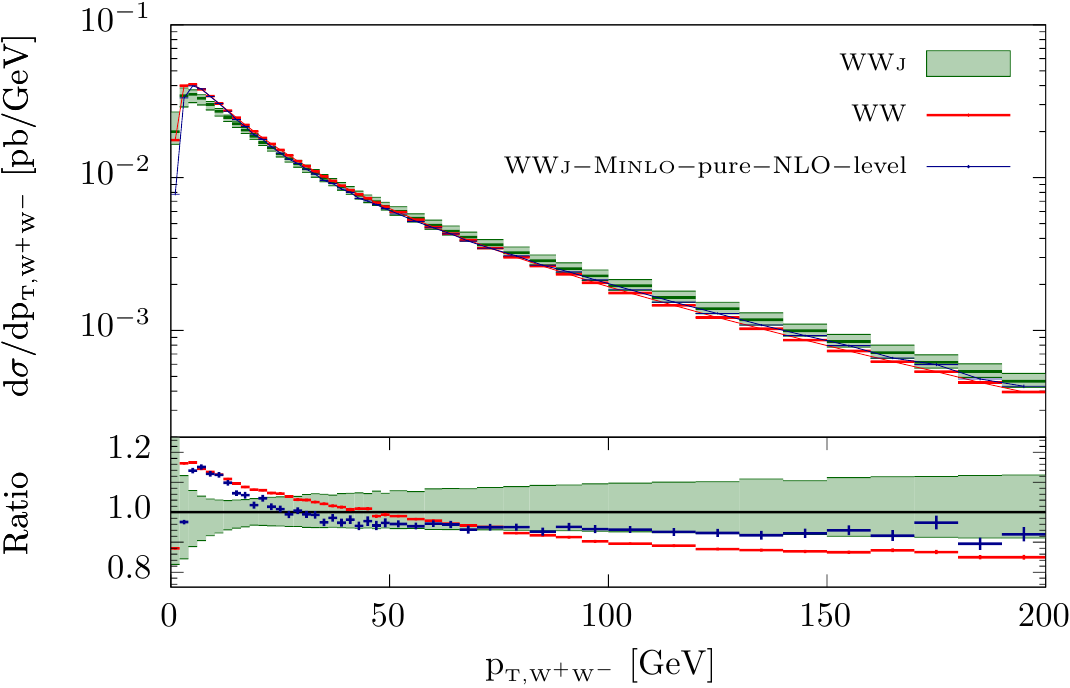}\protect\caption{
    Transverse momentum of the $W^{\scriptscriptstyle{+}}W^{\scriptscriptstyle{-}}$ system as
    predicted by the \noun{WW} (red), \noun{WWj-Minlo} (dark green)
    generators and \noun{WWj-Minlo} at pure NLO level (blue). The simulation of parton shower and hadronization effects
    (but not MPI) is included with  \PYTHIAEIGHT{}.}
\label{fig:wwpt} 
\end{figure}

Next we show in fig.~\ref{fig:wwpt} 
the transverse momentum of the \ww system. At fixed order, predictions
for this observable become divergent as $\ptww$ approaches zero. All
results shown here instead show a physical Sudakov peak and damping at
small transverse momenta, owing to their all-orders resummation of
soft/collinear emission effects; primarily via the \POWHEG{} Sudakov
form factor in the case of the \WW{} generator (red), and through the
\MINLOp Sudakov form factor and scale assignments in \WWJMINLO{}
without (blue) or with parton shower effects (red). In contrast to
fig.~\ref{fig:mw}, the differences between the predictions of the WW
(red) and \noun{WWj-Minlo} (green) generators here are not flat as a
function of $\ptww$, and they are larger in magnitude, both at high
and at low transverse momentum.

At high transverse momenta ($\ptww \sim 100$ GeV), only the
\WWJMINLO{} description is NLO accurate. In this region we observe
that the \WWJMINLO{} spectrum is harder than that of the \WW{}
generator and we interpret the modest $\sim 15\,\%$ difference as
being predominantly due to genuine NLO QCD corrections to the $\ptww$
spectrum. In the low transverse momentum region the WW and \WWJMINLO{}
generators are seen to deviate from one another by up to 15\% in the
Sudakov peak region. On one hand we might well anticipate differences
of roughly this size based on the fact that the WW and
\noun{WWj-Minlo} generators should be expected to differ at roughly
the level of NLL and $\rm{N^{3}LL}_{\sigma}$ terms. On the other hand,
naively at least, it is difficult to reconcile this large difference
in the peak region with the fact that the two codes agree on the total
inclusive cross section to within 3\%. Further investigation shows
that in fact the bare NLO \noun{WWj-Minlo} $\ptww$ spectrum (blue
line) actually follows that of the \noun{Powheg} WW generator
remarkably well in the Sudakov peak region, with the two agreeing to
better than 4-5\% across in the region $\ptww < 60 \rm{GeV}$. Noting
this fact it becomes much easier to understand how the total inclusive
cross sections can be in such good agreement, despite the apparently
sizable differences in the Sudakov peak region. Essentially, the
prediction of the bare \noun{WWj-Minlo} is subsequently acted on in
the simulation chain, first by the \noun{Powheg} hardest radiation
generator (generating the second hardest radiated parton in the case
of \noun{WWj-Minlo}), and then by the parton shower. Both of these
operations exactly respect the unitarity of the cross section, neither
creating, deleting, or reweighting events, however, they will act to
redistribute that cross section through the phase space, albeit
consistently with NLO accuracy. Here, in particular, in the low
$\ptww$ region these multiple emission corrections have the effect of
`smearing out' the more peaked distribution from the bare NLO
\noun{WWj-Minlo} calculation (which, again, tracks closely that of
WW), yielding the more blunt peak of the \noun{WWj-Minlo} generator
seen in the plots.  As a final remark on this aspect, we note also
that these deviations at low $\ptww$ correlate closely with similar
ones in the transverse momentum spectra of the individual $W$ bosons
themselves, for obvious reasons.
We observe that the scale-uncertainty band of the WW code is once
again smaller than the corresponding uncertainty for the
\noun{WWj-Minlo} code. As discussed above, the latter code gives a
more reliable estimate of the size of higher-order corrections which
are not accounted for in our prediction.
Remarkably similar trends to those shown here can be seen in the
comparisons of the \noun{Powheg} W and Z codes to \noun{Wj-Minlo} and
\noun{Zj-Minlo} respectively, for the $W$ and $Z$
$p_{\scriptscriptstyle{\rm{T}}}$ spectra, in
ref.~\cite{Hamilton:2012rf}.

\begin{figure}[htbp]
  \centering{}\includegraphics[scale=0.68]{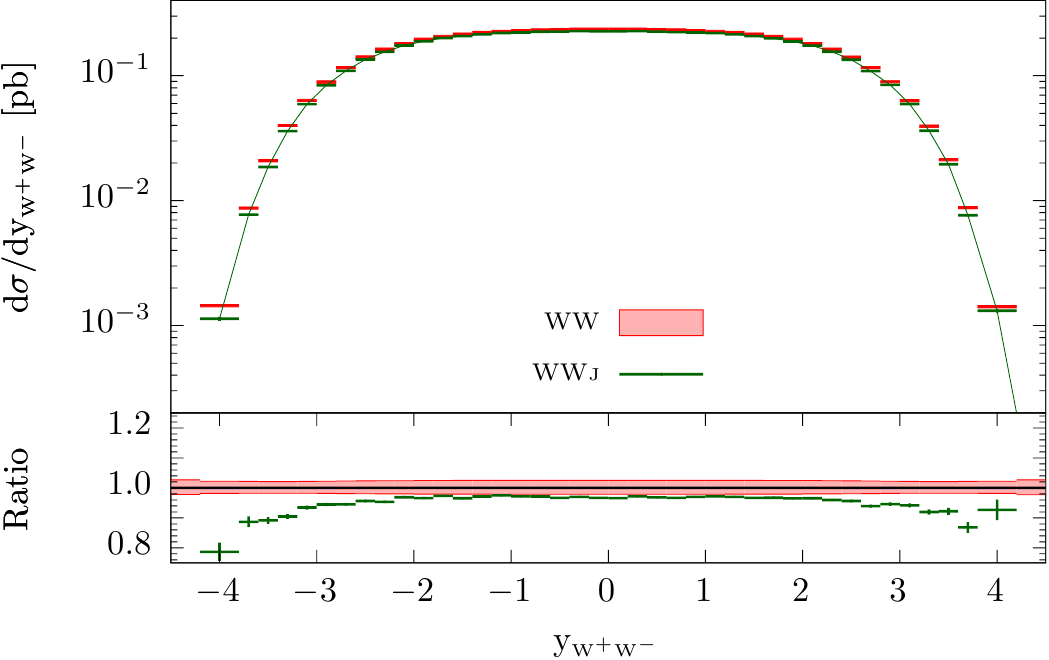}\hspace{0.1cm}\includegraphics[scale=0.68]{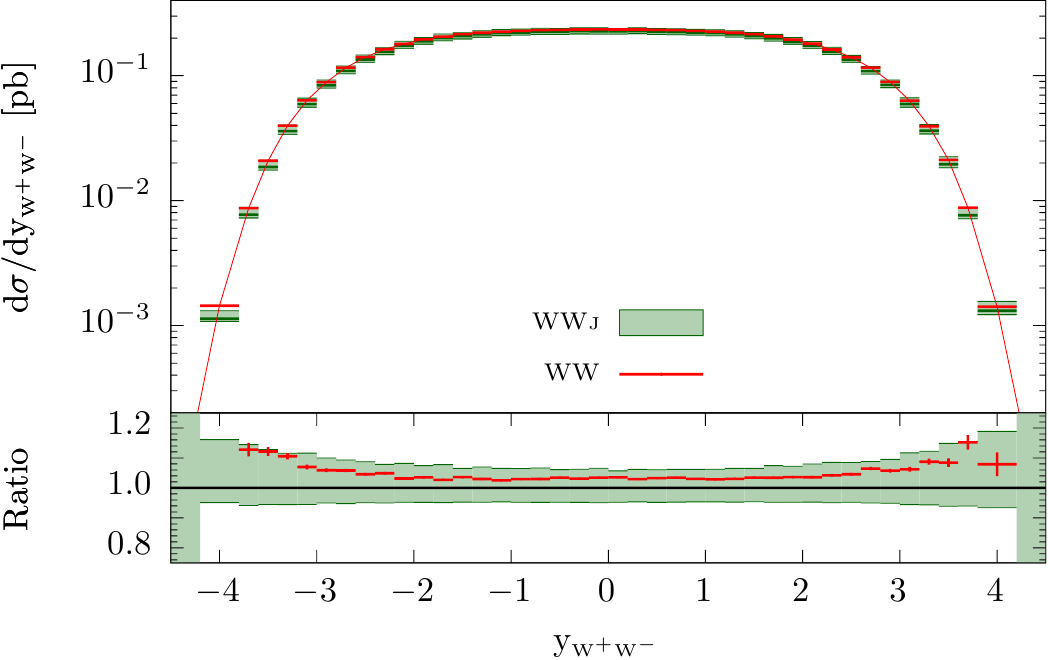}\protect\caption{
    Rapidity of the $W^{\scriptscriptstyle{+}}W^{\scriptscriptstyle{-}}$ system as predicted by
    the \noun{WW} (red) and \noun{WWj-Minlo} (dark green)
    generators. }
\label{fig:wwy} 
\end{figure}

In fig.~\ref{fig:wwy} we show
the rapidity of the
$W^{\scriptscriptstyle{+}}W^{\scriptscriptstyle{-}}$ pair.  As in the
case of the $W^{\scriptscriptstyle{+}}$ mass distribution
(fig.~\ref{fig:mw}), the WW central prediction lies within the
uncertainty of the \WWJMINLO{} generator. On the other hand, at high
rapidities the \WWJMINLO{} predictions are lower than the \WW{}
ones. Here again, the pattern of differences is quantitatively similar
to that found in comparing Z and \noun{Zj-Minlo} predictions, for the
$Z$ rapidity spectrum in the \MINLOp implementations of
ref.~\cite{Hamilton:2012rf}. We add that the high-rapidity regions
here, proportionally, contain more low $\ptww$ events than the central
domain. Thus, we suggest that the deviations seen at high rapidities,
between the \WW{} and \WWJMINLO{} predictions, are strongly correlated
with the comparable deviations in the $\ptww$ spectrum of
fig.~\ref{fig:wwpt}.  Similar behaviour is found for the rapidity
distributions of the individual $W^{\scriptscriptstyle{\pm}}$ bosons,
as well as their decay products.

The missing transverse momentum ($\ptmiss$) distribution is shown in
\begin{figure}[htbp]
  \centering{}\includegraphics[scale=0.68]{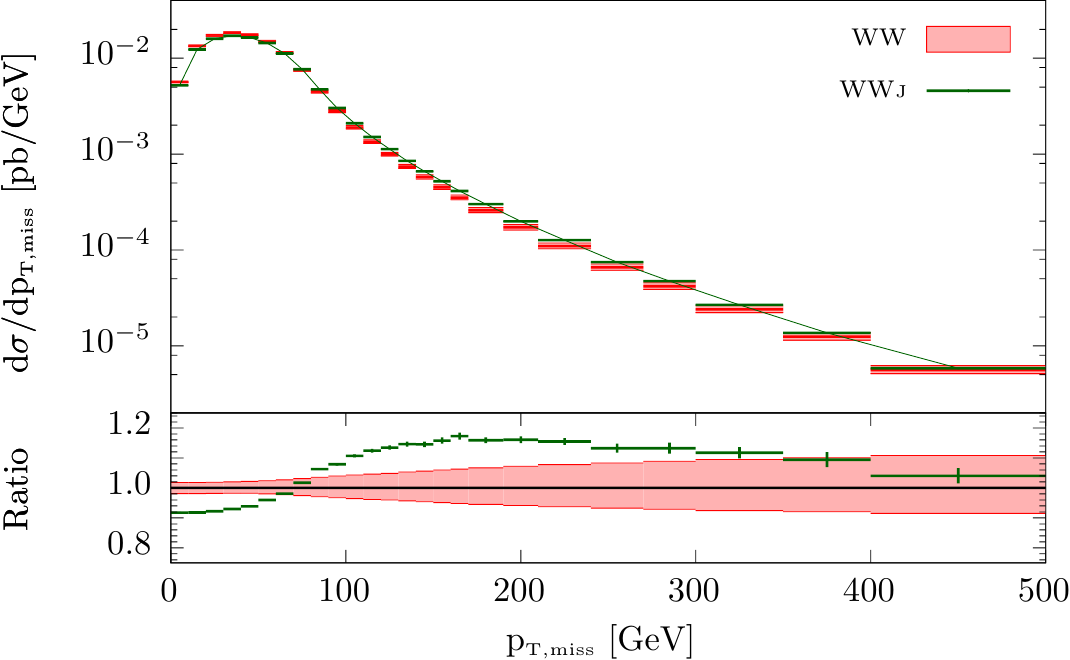}\hspace{0.1cm}\includegraphics[scale=0.68]{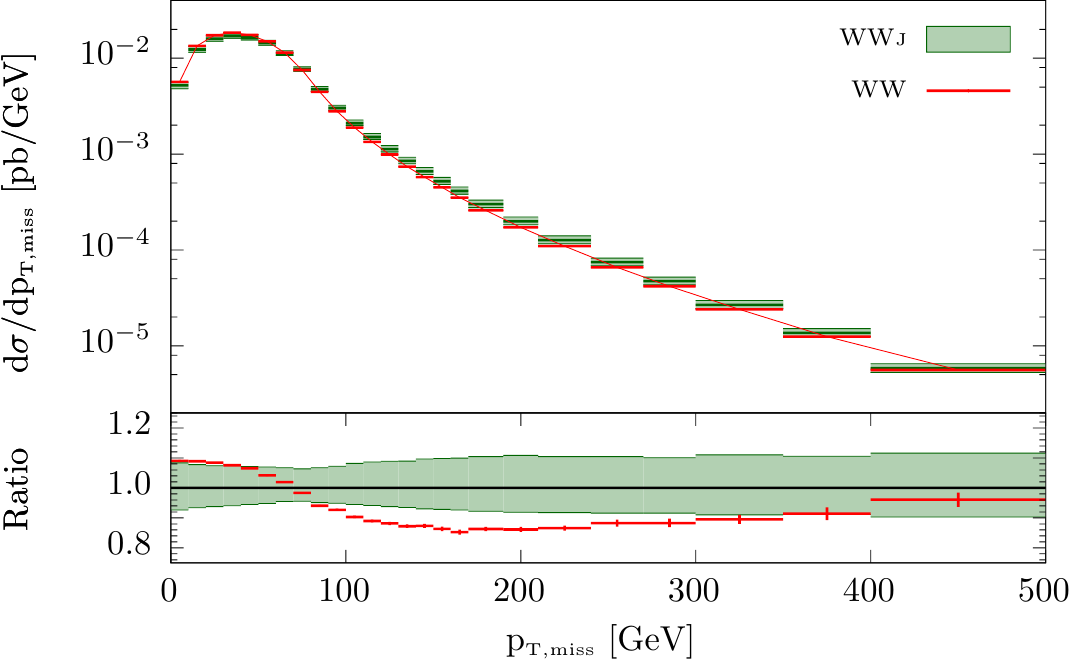}\protect\caption{\label{fig:ptmiss}
    Missing transverse momentum as predicted by the \noun{WW} (red)
    and \noun{WWj-Minlo} (dark green) generators. }
\end{figure}
fig.~\ref{fig:ptmiss}. Not surprisingly, this observable shows a
pattern qualitatively similar to the one observed in
fig.~\ref{fig:wwpt} for the transverse momentum of the
$W^{\scriptscriptstyle{+}}W^{\scriptscriptstyle{-}}$ system.  Although
the $W^{\scriptscriptstyle{+}}W^{\scriptscriptstyle{-}}$ transverse
momentum is shared between leptons and neutrinos, one expects that the
large $\ptmiss$ tail is mostly populated by events where the
$W^{\scriptscriptstyle{+}}W^{\scriptscriptstyle{-}}$ system had a
large boost, hence the \WWJMINLO{} result displays a cross section
larger than the \WW{} one.  The large differences observed at small
values of $\ptww$ get instead partly diluted when looking at
$\ptmiss$. This can be understood by considering the underlying weak
decay $W^{\scriptscriptstyle{+}} W^{\scriptscriptstyle{-}} \to l^+ l^-
\nu \bar{\nu}$: even when the
$W^{\scriptscriptstyle{+}}W^{\scriptscriptstyle{-}}$ system is almost
at rest, the transverse momentum of each neutrino is of order
$m_{{\scriptscriptstyle W}}/2$ or more (depending on how boosted the
$W$ boson is off which it is emitted). This consideration, together
with the fact that the missing energy is the absolute value of a
vectorial sum of two transverse momenta, justifies why differences
that at the peak of the $\ptww$ distribution reached a factor 1.16
become averaged out in the $\ptmiss$ spectrum, which, at low values,
only exhibits 10\% differences at most between the \WW{} and
\WWJMINLO{} simulations.

\subsubsection{Jet associated production \label{sub:Comparison-of-jet-associated-production}}

We now turn to results where at least one jet is required in the final
state. Jets are reconstructed using the anti-$k_t$
algorithm~\cite{Cacciari:2008gp} as implemented in
FastJet~\cite{Cacciari:2011ma}. In the following, we have chosen
$R=0.4$ and jets are required to have $\ptj > \ptmin = 25$ GeV.

\begin{figure}[htbp]
  \centering{}\includegraphics[scale=0.68]{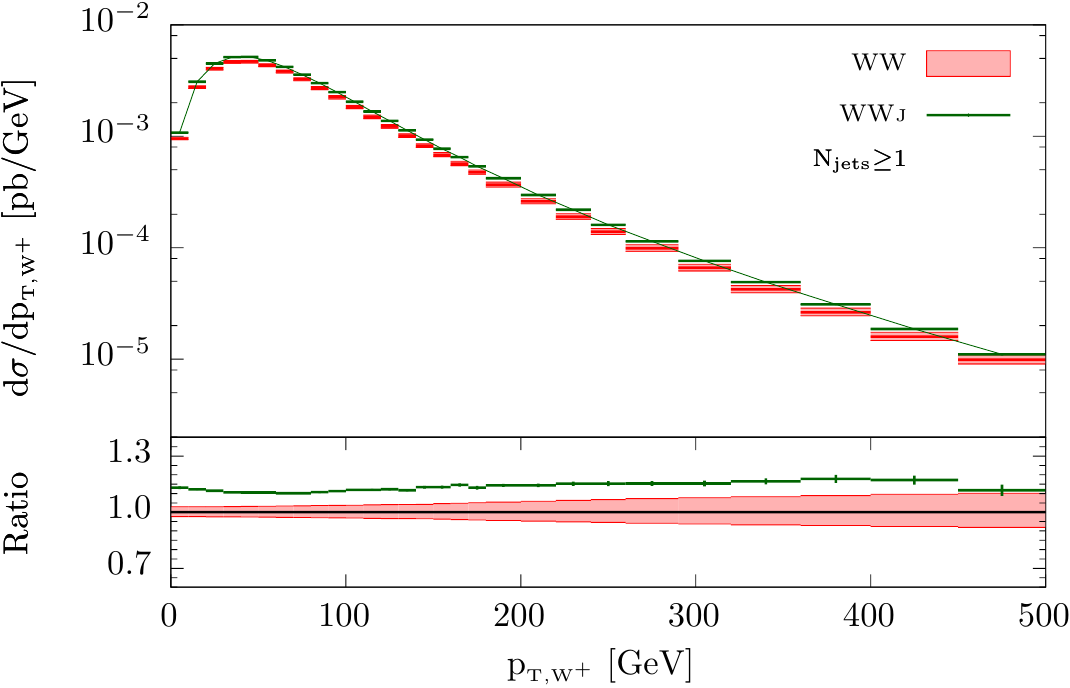}\hspace{0.1cm}\includegraphics[scale=0.68]{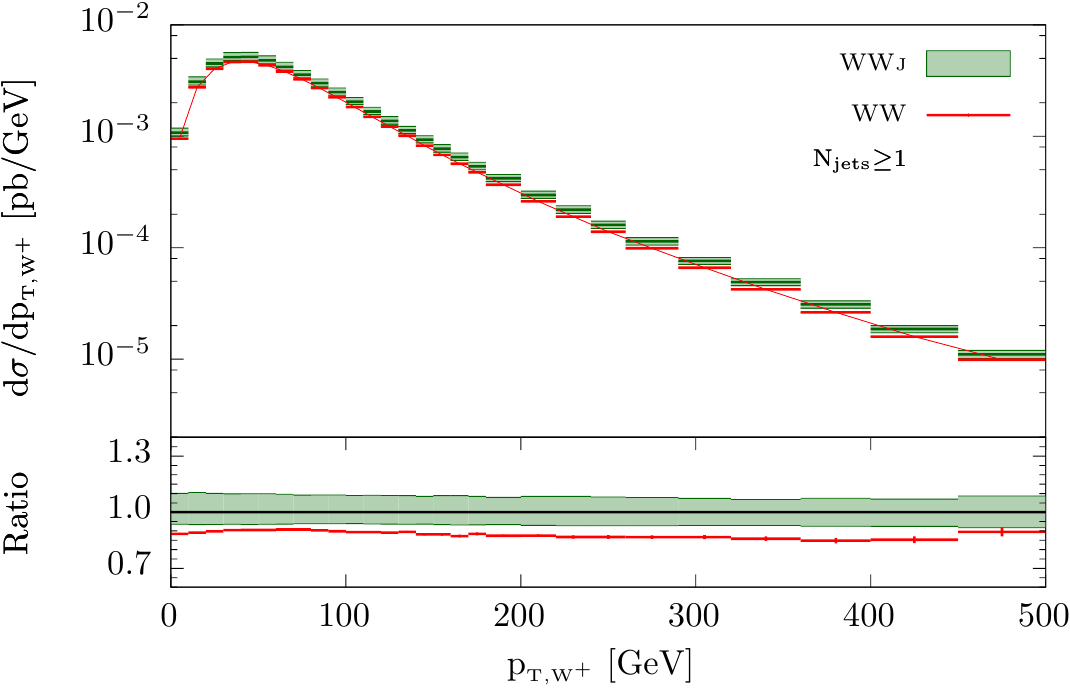}\protect\caption{\label{fig:ptw_1j}
    The $W^{\scriptscriptstyle{+}}$ boson transverse momentum in the $1$-jet
    region, as predicted by the \noun{WW} (red) and \noun{WWj-Minlo} (dark
    green) generators.}
\end{figure}

In fig.~\ref{fig:ptw_1j} we show the $W^{\scriptscriptstyle{+}}$
transverse momentum distribution in events with at least one jet.
This distribution is described at NLO accuracy by both generators when
no jet-cut is imposed. However, when we require to have one jet in the
final state, the \WW{} generator is only LO accurate. In this case,
due to the inclusive nature of this observable with respect to extra
QCD radiation, NLO corrections amount to an overall $K$ factor, and
hence we find good agreement in the shape of the two distributions.

\begin{figure}[htbp]
  \centering{}\includegraphics[scale=0.68]{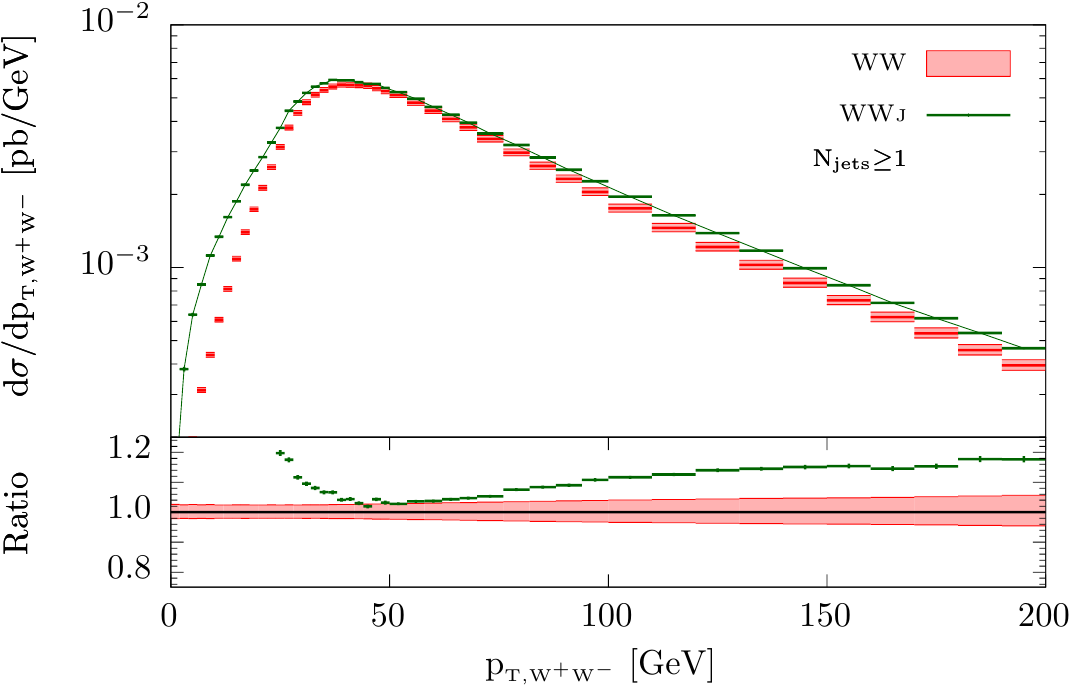}\hspace{0.1cm}\includegraphics[scale=0.68]{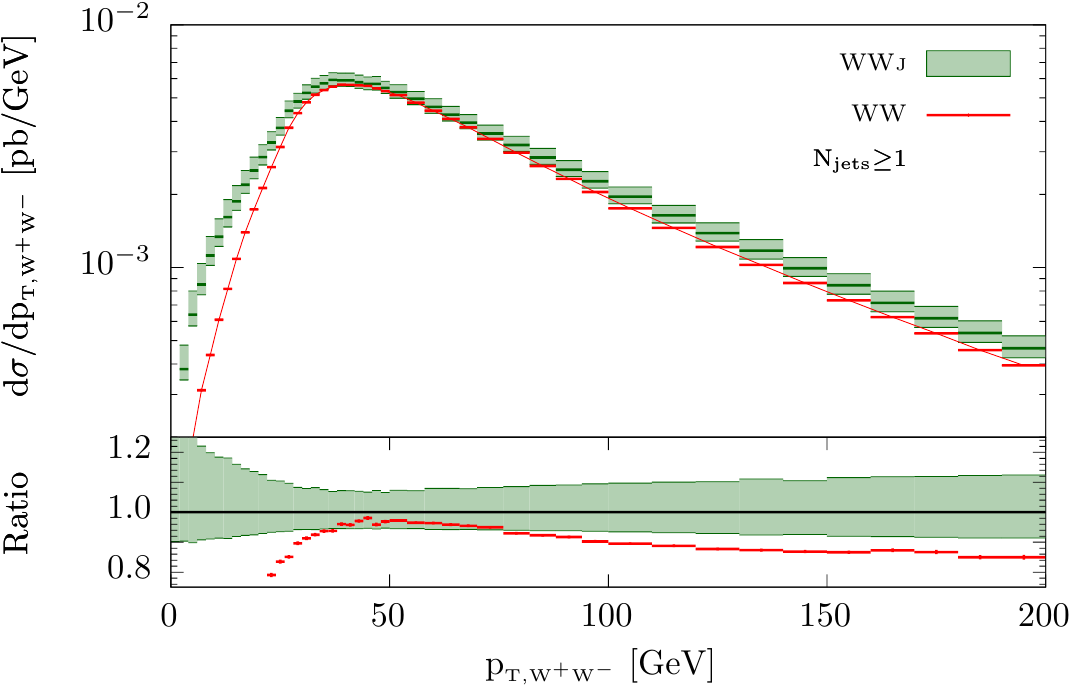}\protect\caption{\label{fig:ptww_1j}
    The $W^{\scriptscriptstyle{+}}W^{\scriptscriptstyle{-}}$ transverse momentum
    in the $1$-jet region, as predicted by the \noun{WW} (red) and 
    \noun{WWj-Minlo} (dark green) generators.}
\end{figure}

In fig.~\ref{fig:ptww_1j} we now examine the transverse momentum
distribution of the
$W^{\scriptscriptstyle{+}}W^{\scriptscriptstyle{-}}$ system in events
with at least one jet.
In this case, when the transverse momentum is large ($\ptww \gg
\ptmin$) we recover the behaviour observed in fig.~\ref{fig:wwpt}. On
the other hand we observe a very different behaviour at small $\ptww$.
Given the jet requirement, the region $0<\ptww < 25 $ GeV is populated
only by events with at least two QCD emissions. Therefore, for the
\WW{} generator this region is populated by the parton shower
only. For the \WWJMINLO{} generator, on the other hand, two-parton
configurations are also provided by leading order matrix elements;
these tend to populate this region more strongly than the shower alone
since they tend to provide harder QCD radiation against which the
first jet can recoil, leaving room for $\ptww$ to be smaller. The
shape change that we observe for this distribution at $\ptww = 25$ GeV
has to do with the \WWJMINLO{} code switching from being LO accurate
below that threshold to NLO accurate above it. The LO behaviour of the
\noun{WWj-Minlo} generator in this region is also evident from the
widening of the \WWJMINLO{} uncertainty band below $\ptww = 25$ GeV.

\begin{figure}[htbp]
  \centering{}\includegraphics[scale=0.68]{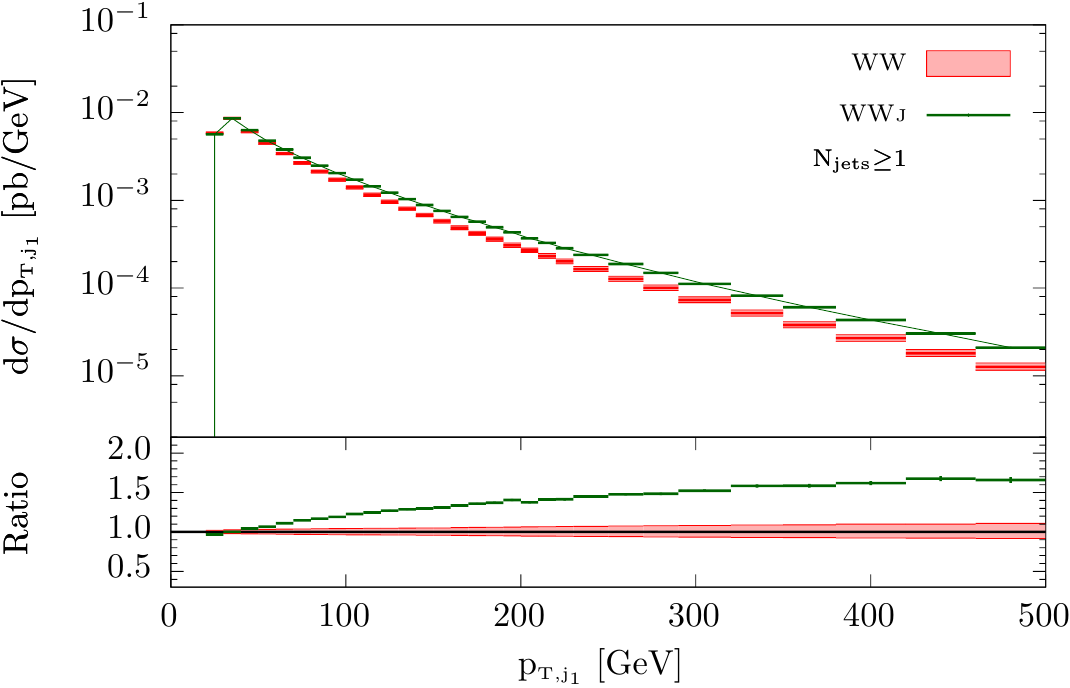}\hspace{0.1cm}\includegraphics[scale=0.68]{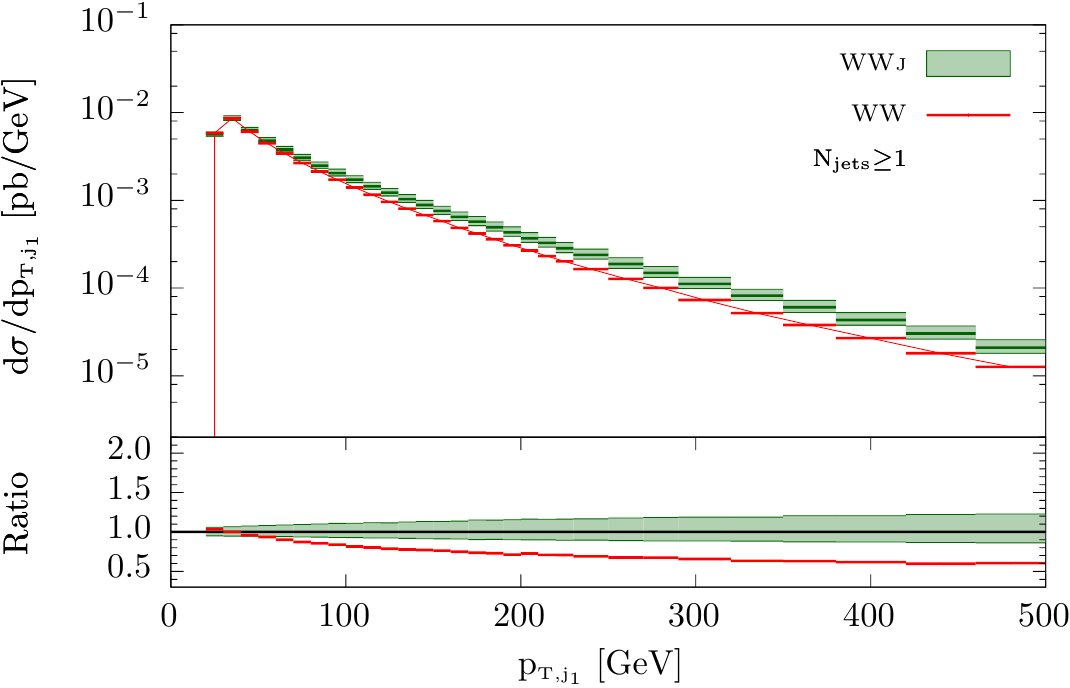}\protect
  \caption{\label{fig:ptj1}Leading jet transverse momentum as predicted by the \noun{WW}
    (red) and \noun{WWj-Minlo} (dark green) generators. }
\end{figure}

In fig.~\ref{fig:ptj1} we show
the transverse momentum spectrum of the hardest jet. This is a
quantity that is described only at LO level by the \noun{WW}
generator, whereas NLO corrections are included in the
\noun{WWj-Minlo} program. We notice that the two results start to
deviate at fairly modest transverse momenta, and differences up to a
factor of almost 2 can be noticed in the tail. This is due to the
absence of radiative corrections in the \WW{} generator. In particular
we have verified that in the tail region the contribution from events
containing 2 well separated partons is sizable. The \WW{} program
underestimates their rates (as shown also in the next plot), since it
doesn't contain the corresponding exact matrix elements.  The
uncertainty band of the \WWJMINLO{} result shows the size and pattern
that one would expect to see in a NLO-accurate prediction. On the
other hand, especially in the low-to-medium range, the uncertainty
band for the \WW{} result is thinner then the \WWJMINLO{} one, despite
the nominal accuracy of the \WW{} generator is just LO for this
observable. This is explained as follows: in this region the scale
variation for the \WW{} predictions is due to the scale dependence of
the $\bar{B}$ function, which is a NLO-accurate quantity. At larger
$\pt$ values, where \texttt{bornzerodamp} is active, the band slowly
thickens, giving an uncertainty of $\pm 10\%$.

Another interesting observable to consider is
\begin{figure}[htbp]
  \centering{}\includegraphics[scale=0.68]{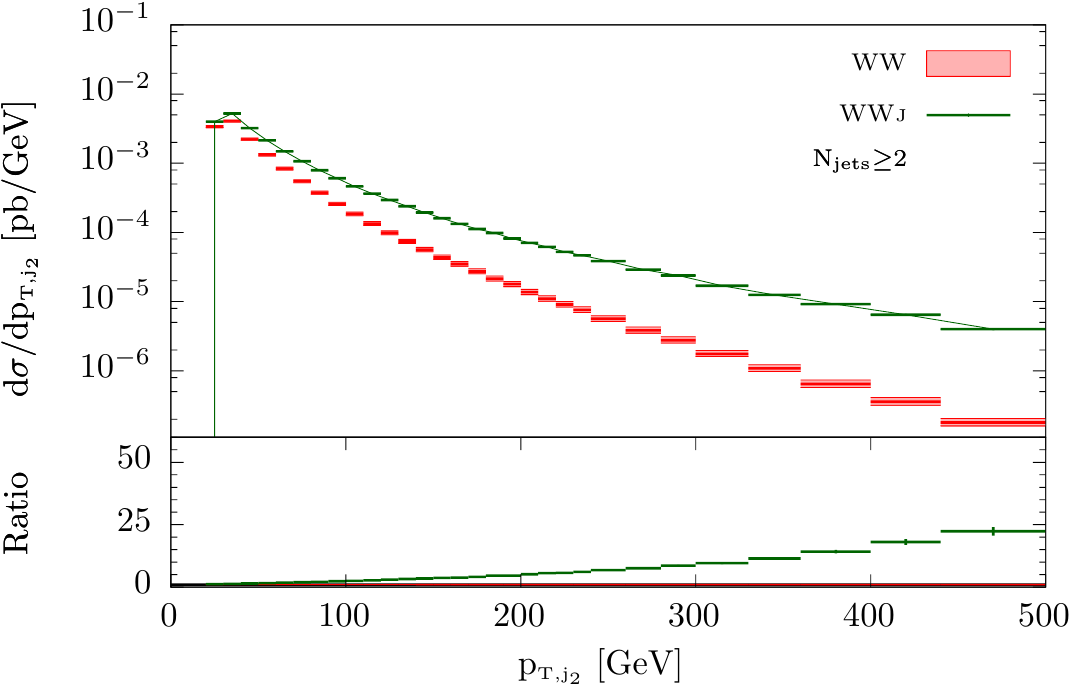}\hspace{0.1cm}\includegraphics[scale=0.68]{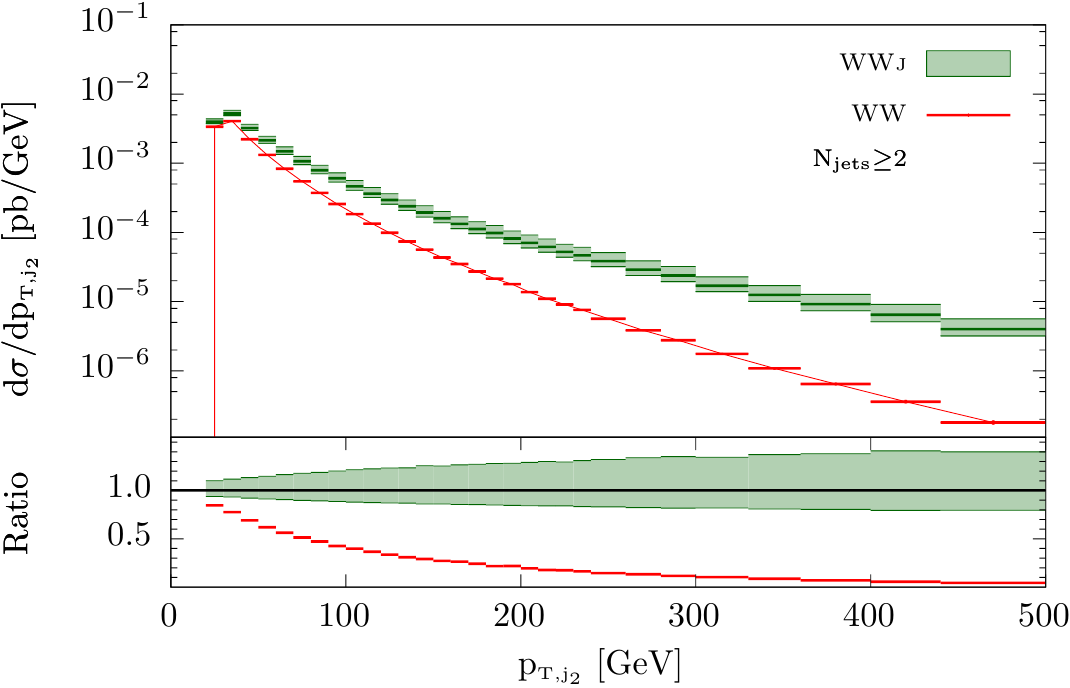}\protect\caption{\label{fig:ptj2}
    Second jet transverse momentum as predicted by the \noun{WW} (red)
    and \noun{WWj-Minlo} (dark green) generators. }
\end{figure}
the transverse momentum spectrum of the second hardest jet, which we
show in fig.~\ref{fig:ptj2}. Not surprisingly, here we observe huge
differences among the two generators. In the \WW{} code only the
hardest radiation is generated by \POWHEG{}. Hence the particles that
constitute the second jet are only produced via parton showering: when
large $\ptjj$ regions are probed, the \WW{} code is bound to predict
an unreliable cross section (too small in this case). The \WWJMINLO{}
prediction is instead more accurate, since the matrix elements
describing the production of two separated outgoing partons are
included exactly, although only at LO.  The LO nature of this result
is reflected in the relatively large uncertainty band.

\begin{figure}[htbp]
  \centering{}\includegraphics[scale=0.68]{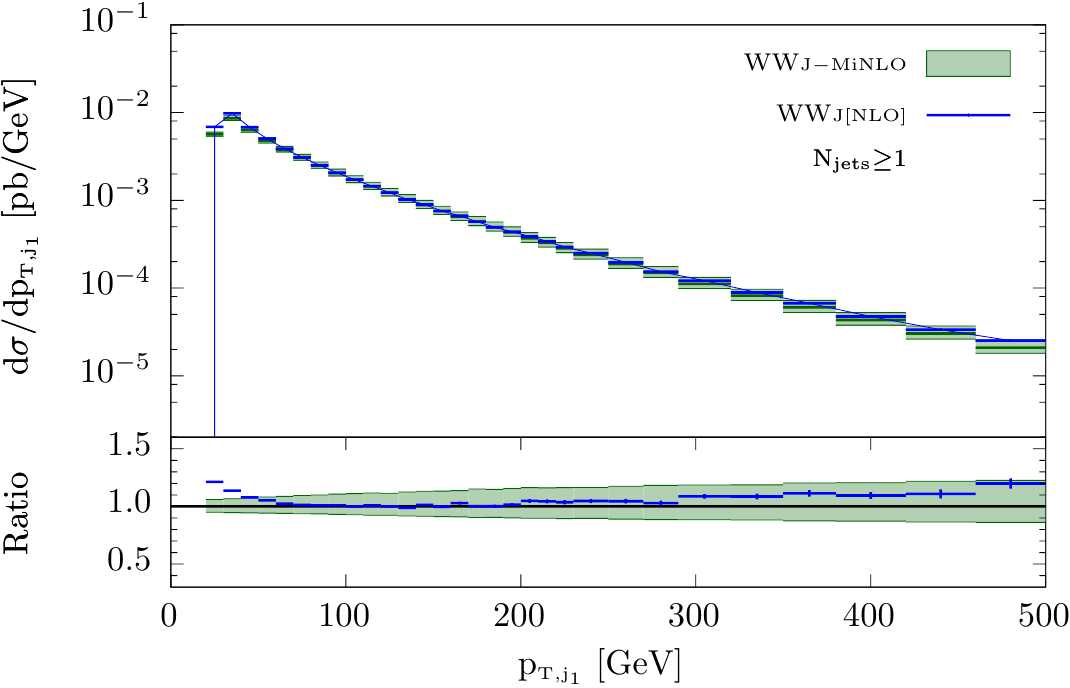}\hspace{0.1cm}\includegraphics[scale=0.68]{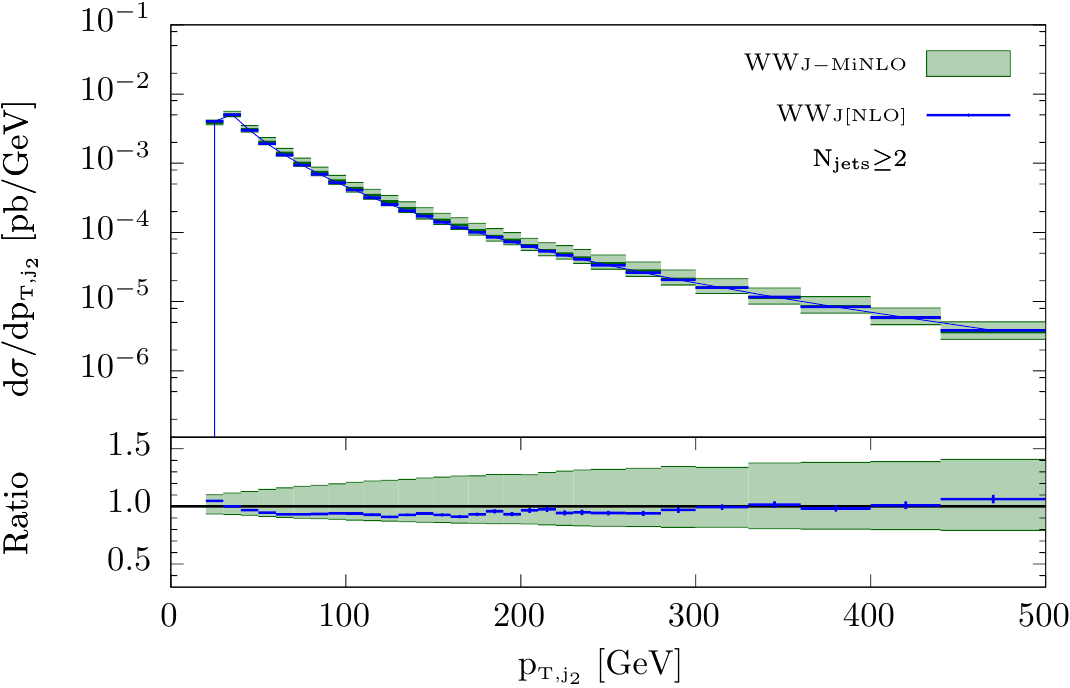}\protect\caption{\label{fig:ptjets_nlo}
    First and second jet transverse momentum as predicted by the
    \noun{WWj-Minlo} (dark green) generators compared against a
    fixed-order NLO computation for the same process (\noun{WWj
      [NLO]}, blue).}
\end{figure}

After having shown how the \WWJMINLO{} generator compares against the
\WW{} one for jet observables, we find it useful to compare, for the
same observables, \WWJMINLO{} against a NLO computation (without any
\MINLO{} improvement) for the process $pp\to W^+W^-j$.\footnote{This
  result was obtained running at fixed-order the \WWJMINLO{} code,
  switching off the \MINLO{} machinery but including a 10 GeV
  generation cut for the hardest parton transverse
  momentum. Renormalization and factorization scales have been set
  equal to $2m_{{\scriptscriptstyle W}}$.} These comparisons are shown
in fig.~\ref{fig:ptjets_nlo}.
In the left panel we observe that the \WWJMINLO{} prediction for the
hardest jet $\pt$ spectrum agrees very well with the fixed-order
result for the transverse momentum spectrum of the hardest jet. The
moderate differences close to the threshold to produce one jet are
likely due to the use of different scales as well as to the presence
of the \MINLO{} Sudakov. At larger values of $\ptj$ the two
predictions are compatible, although the NLO result exhibits a
slightly harder spectrum. This is easily explained by recalling that,
at large transverse momenta, although the \MINLO{} Sudakov switches
off, the \WWJMINLO{} scale prescription is used: the \WWJMINLO{} line
is obtained with a dynamical scale choice,~i.e. a scale choice
certainly larger than the one used for the NLO computation, where we
have chosen $\mu=2 m_{{\scriptscriptstyle W}}$. This consideration is
also supported by the fact that the NLO and \WWJMINLO{} predictions
agree extremely well for $\ptj\simeq 2 m_{{\scriptscriptstyle W}}$.
In the right panel of fig.~\ref{fig:ptjets_nlo} we show instead the
comparison for the second hardest jet, which agree quite well.

\section{Conclusion and outlook\label{sec:Conclusion}}

In this paper we have extended the \MINLOp method to merge
$W^{\scriptscriptstyle{+}}W^{\scriptscriptstyle{-}}$ and
$W^{\scriptscriptstyle{+}}W^{\scriptscriptstyle{-}}\!$+jet NLO
calculations, preserving NLO accuracy throughout the 0- and 1-jet
phase space.  The merging is achieved without the introduction of an
unphysical merging scale; such a scale formally spoils the NLO
accuracy of merged samples in some regions of phase space.

The work presented here exemplifies the extension of the \MINLOp
method, in its original form, to general colour singlet production
processes, where the virtual corrections to the lowest multiplicity
process are non-trivial.
The method relies on the fact that the $\textrm{NNLL}$ transverse
momentum resummation coefficient $B_2$ is the same in
$W^{\scriptscriptstyle{+}}W^{\scriptscriptstyle{-}}$ and Drell-Yan
production, save for a process-dependent term proportional to the
virtual corrections affecting the leading order process. By carefully
replacing this process-dependent component in the \MINLOp recipe for
Drell-Yan type reactions, with the analogous virtual correction to
$W^{\scriptscriptstyle{+}}W^{\scriptscriptstyle{-}}$ production, the
\MINLOp method becomes directly applicable to the latter. The same
replacement procedure holds for extending \MINLOp to all colour
singlet production
processes. 
 
We have performed an extensive phenomenological study, comparing our
results to conventional
$W^{\scriptscriptstyle{+}}W^{\scriptscriptstyle{-}}$ \noun{Nlops} and
$W^{\scriptscriptstyle{+}}W^{\scriptscriptstyle{-}}$+jet NLO
calculations. In general we find good agreement, with standard NLO
results lying within the uncertainty band of the \MINLOp improved
prediction. On the other hand, the scale uncertainty of the \MINLOp
results for inclusive observables, while being less than 7\%, is
larger than that of conventional NLO, by a factor of 2-3.

Although it has been proven that the \MINLOp method yields NLO
accuracy for 0-jet and inclusive quantities, and that scale variations
therein give rise to only NNLO-sized shifts, the precise value of
these relative
${\cal{O}}(\alpha_{\scriptscriptstyle{\mathrm{S}}}^{2})$ ambiguities
is understood to be formally different in each case.  The propagation
of scale variation in the \MINLOp approach is complicated, however,
it's somewhat natural to expect that \MINLOp should tend to exhibit a
larger scale uncertainty than conventional NLO calculations: in the
former, the scale for evaluating the strong coupling constant and PDFs
is mandatorily $\ptww$, thus, scale variation for the great majority
of events forming the inclusive cross section, in the vicinity of the
Sudakov peak, takes place around significantly smaller values in
\MINLOp than in conventional NLO computations
($\mu_{{\scriptscriptstyle R}} \sim \mu_{{\scriptscriptstyle F}}
\gtrsim {\cal{O}}(m_{{\scriptscriptstyle W}})$).  Considering the
differences reported between NLO and NNLO results in the recent study
of ref.~\cite{Grazzini:2016ctr}, we regard the theoretical uncertainty
estimate from our \MINLOp computation as being quite reasonable, while
that of conventional NLO is a poor underestimate.

In conclusion we find that the \WWJMINLO{} generator supersedes the
\WW{} generator, retaining the NLO accuracy of the latter for
inclusive quantities and augmenting it with a more
reliable, NLO accurate description of hard radiation. Furthermore, for
fully inclusive, 0-jet, and 1-jet quantities the \WWJMINLO{} generator
gives a more realistic estimate of the theoretical uncertainty.

Finally, we remark that with a \MINLOp accurate simulation at hand, it
becomes straightforward, theoretically, to upgrade it, through a
reweighting procedure, to obtain \textsc{NNLOPS} accurate
predictions. In the present case, taking into account the
$W^{\scriptscriptstyle{+}}W^{\scriptscriptstyle{-}}$ decays, the high
dimensionality of the Born phase space makes this task far from
trivial in practice. We leave this to future work.

\section*{Acknowledgements}

We would like to thank G.~Luisoni for his assistance with the code
\noun{GoSam}.  The research of GZ is supported by the ERC consolidator
grant 614577 {\it HICCUP -- High Impact Cross Section Calculations for
  Ultimate Precision}. PM and ER have also benefited from the ERC
grant {\it HICCUP}.  TM is supported by U.S. DOE grant
DE-AC02-05CH11231 and acknowledges computational resources provided
through ERC grant number 291377 {\it LHCtheory}.  PM was partly
supported by the Swiss National Science Foundation (SNF) under grant
PBZHP2-147297.
\noindent We all gratefully acknowledge the Mainz Institute for
Theoretical Physics (MITP) for its hospitality and support while part
of this work was carried out. Additionally, GZ gratefully acknowledges
KITP and PM and ER acknowledge the CERN's Theory Department for
hospitality while completing this work.

\bibliographystyle{jhep}
\bibliography{paper}

\providecommand{\href}[2]{#2}\begingroup\raggedright\begin{thebibliography}{10}

\bibitem{Aaltonen:2009aa}
{\bf CDF} Collaboration, T.~Aaltonen et~al., {\it {Measurement of the $W^+ W^-$
  Production Cross Section and Search for Anomalous $WW \gamma$ and $WWZ$
  Couplings in $p \bar{p}$ Collisions at $\sqrt{s}=1.96$~TeV}},  {\em Phys.
  Rev. Lett.} {\bf 104} (2010) 201801,
  [\href{http://xxx.lanl.gov/abs/0912.4500}{{\tt arXiv:0912.4500}}].

\bibitem{CDFnote}
{\bf CDF} Collaboration, T.~Aaltonen et~al., {\it {Measurement of the $W^+ W^-$
  Production Cross Section and Differential Cross Section with Jets in $p
  \bar{p}$ Collisions at $\sqrt{s}=1.96$~TeV}},
  \href{http://xxx.lanl.gov/abs/http://www-cdf.fnal.gov/physics/ewk/2014/WWjets/cdf11098\_WW+jets.pdf}{{\tt
  http://www-cdf.fnal.gov/physics/ewk/2014/WWjets/cdf11098\_WW+jets.pdf}}.

\bibitem{Abazov:2012ze}
{\bf D0} Collaboration, V.~M. Abazov et~al., {\it {Limits on anomalous
  trilinear gauge boson couplings from $WW$, $WZ$ and $W\gamma$ production in
  $p\bar{p}$ collisions at $\sqrt{s}=1.96$ TeV}},  {\em Phys. Lett.} {\bf B718}
  (2012) 451--459, [\href{http://xxx.lanl.gov/abs/1208.5458}{{\tt
  arXiv:1208.5458}}].

\bibitem{Aad:2016wpd}
{\bf ATLAS} Collaboration, G.~Aad et~al., {\it {Measurement of total and
  differential $W^+W^-$ production cross sections in proton-proton collisions
  at $\sqrt{s}=$ 8 TeV with the ATLAS detector and limits on anomalous
  triple-gauge-boson couplings}},
  \href{http://xxx.lanl.gov/abs/1603.0170}{{\tt arXiv:1603.0170}}.

\bibitem{Khachatryan:2015sga}
{\bf CMS} Collaboration, V.~Khachatryan et~al., {\it {Measurement of the
  $W^+W^-$ cross section in pp collisions at $\sqrt{s}$ = 8 TeV and limits on
  anomalous gauge couplings}},  \href{http://xxx.lanl.gov/abs/1507.0326}{{\tt
  arXiv:1507.0326}}.

\bibitem{Brown:1978mq}
R.~W. Brown and K.~O. Mikaelian, {\it {W+ W- and Z0 Z0 Pair Production in e+
  e-, p p, p anti-p Colliding Beams}},  {\em Phys. Rev.} {\bf D19} (1979) 922.

\bibitem{Dixon:1998py}
L.~J. Dixon, Z.~Kunszt, and A.~Signer, {\it {Helicity amplitudes for O(alpha-s)
  production of $W^{+} W^{-}$, $W^\pm Z$, $Z Z$, $W^\pm \gamma$, or $Z \gamma$
  pairs at hadron colliders}},  {\em Nucl. Phys.} {\bf B531} (1998) 3--23,
  [\href{http://xxx.lanl.gov/abs/hep-ph/9803250}{{\tt hep-ph/9803250}}].

\bibitem{Campbell:1999ah}
J.~M. Campbell and R.~K. Ellis, {\it {An Update on vector boson pair production
  at hadron colliders}},  {\em Phys. Rev.} {\bf D60} (1999) 113006,
  [\href{http://xxx.lanl.gov/abs/hep-ph/9905386}{{\tt hep-ph/9905386}}].

\bibitem{Dixon:1999di}
L.~J. Dixon, Z.~Kunszt, and A.~Signer, {\it {Vector boson pair production in
  hadronic collisions at order $\alpha_s$ : Lepton correlations and anomalous
  couplings}},  {\em Phys. Rev.} {\bf D60} (1999) 114037,
  [\href{http://xxx.lanl.gov/abs/hep-ph/9907305}{{\tt hep-ph/9907305}}].

\bibitem{Campbell:2011bn}
J.~M. Campbell, R.~K. Ellis, and C.~Williams, {\it {Vector boson pair
  production at the LHC}},  {\em JHEP} {\bf 07} (2011) 018,
  [\href{http://xxx.lanl.gov/abs/1105.0020}{{\tt arXiv:1105.0020}}].

\bibitem{Campbell:2007ev}
J.~M. Campbell, R.~K. Ellis, and G.~Zanderighi, {\it {Next-to-leading order
  predictions for $WW+1$ jet distributions at the LHC}},  {\em JHEP} {\bf 12}
  (2007) 056, [\href{http://xxx.lanl.gov/abs/0710.1832}{{\tt
  arXiv:0710.1832}}].

\bibitem{Melia:2011dw}
T.~Melia, K.~Melnikov, R.~Rontsch, and G.~Zanderighi, {\it {NLO QCD corrections
  for $W^+W^-$ pair production in association with two jets at hadron
  colliders}},  {\em Phys. Rev.} {\bf D83} (2011) 114043,
  [\href{http://xxx.lanl.gov/abs/1104.2327}{{\tt arXiv:1104.2327}}].

\bibitem{Dicus:1987dj}
D.~A. Dicus, C.~Kao, and W.~W. Repko, {\it {Gluon Production of Gauge Bosons}},
   {\em Phys. Rev.} {\bf D36} (1987) 1570.

\bibitem{Glover:1988fe}
E.~W.~N. Glover and J.~J. van~der Bij, {\it {Vector boson pair production via
  gluon fusion}},  {\em Phys. Lett.} {\bf B219} (1989) 488.

\bibitem{Binoth:2005ua}
T.~Binoth, M.~Ciccolini, N.~Kauer, and M.~Kramer, {\it {Gluon-induced WW
  background to Higgs boson searches at the LHC}},  {\em JHEP} {\bf 03} (2005)
  065, [\href{http://xxx.lanl.gov/abs/hep-ph/0503094}{{\tt hep-ph/0503094}}].

\bibitem{Binoth:2006mf}
T.~Binoth, M.~Ciccolini, N.~Kauer, and M.~Kramer, {\it {Gluon-induced W-boson
  pair production at the LHC}},  {\em JHEP} {\bf 12} (2006) 046,
  [\href{http://xxx.lanl.gov/abs/hep-ph/0611170}{{\tt hep-ph/0611170}}].

\bibitem{Campbell:2011cu}
J.~M. Campbell, R.~K. Ellis, and C.~Williams, {\it {Gluon-Gluon Contributions
  to W+ W- Production and Higgs Interference Effects}},  {\em JHEP} {\bf 10}
  (2011) 005, [\href{http://xxx.lanl.gov/abs/1107.5569}{{\tt
  arXiv:1107.5569}}].

\bibitem{Melia:2012zg}
T.~Melia, K.~Melnikov, R.~Rontsch, M.~Schulze, and G.~Zanderighi, {\it {Gluon
  fusion contribution to W+W- + jet production}},  {\em JHEP} {\bf 08} (2012)
  115, [\href{http://xxx.lanl.gov/abs/1205.6987}{{\tt arXiv:1205.6987}}].

\bibitem{Gehrmann:2014fva}
T.~Gehrmann, M.~Grazzini, S.~Kallweit, P.~Maierh{\"o}fer, A.~von Manteuffel,
  S.~Pozzorini, D.~Rathlev, and L.~Tancredi, {\it {$W^+W^-$ Production at
  Hadron Colliders in Next to Next to Leading Order QCD}},  {\em Phys. Rev.
  Lett.} {\bf 113} (2014), no.~21 212001,
  [\href{http://xxx.lanl.gov/abs/1408.5243}{{\tt arXiv:1408.5243}}].

\bibitem{Gehrmann:2014bfa}
T.~Gehrmann, A.~von Manteuffel, L.~Tancredi, and E.~Weihs, {\it {The two-loop
  master integrals for $q\overline{q} \to VV$}},  {\em JHEP} {\bf 06} (2014)
  032, [\href{http://xxx.lanl.gov/abs/1404.4853}{{\tt arXiv:1404.4853}}].

\bibitem{Caola:2014iua}
F.~Caola, J.~M. Henn, K.~Melnikov, A.~V. Smirnov, and V.~A. Smirnov, {\it
  {Two-loop helicity amplitudes for the production of two off-shell electroweak
  bosons in quark-antiquark collisions}},  {\em JHEP} {\bf 11} (2014) 041,
  [\href{http://xxx.lanl.gov/abs/1408.6409}{{\tt arXiv:1408.6409}}].

\bibitem{Gehrmann:2015ora}
T.~Gehrmann, A.~von Manteuffel, and L.~Tancredi, {\it {The two-loop helicity
  amplitudes for $ q\overline{q}^{\prime}\to {V}_1{V}_2\to 4 $ leptons}},  {\em
  JHEP} {\bf 09} (2015) 128, [\href{http://xxx.lanl.gov/abs/1503.0481}{{\tt
  arXiv:1503.0481}}].

\bibitem{Caola:2015ila}
F.~Caola, J.~M. Henn, K.~Melnikov, A.~V. Smirnov, and V.~A. Smirnov, {\it
  {Two-loop helicity amplitudes for the production of two off-shell electroweak
  bosons in gluon fusion}},  {\em JHEP} {\bf 06} (2015) 129,
  [\href{http://xxx.lanl.gov/abs/1503.0875}{{\tt arXiv:1503.0875}}].

\bibitem{vonManteuffel:2015msa}
A.~von Manteuffel and L.~Tancredi, {\it {The two-loop helicity amplitudes for
  $gg \to V_1 V_2 \to 4~\mathrm{leptons}$}},  {\em JHEP} {\bf 06} (2015) 197,
  [\href{http://xxx.lanl.gov/abs/1503.0883}{{\tt arXiv:1503.0883}}].

\bibitem{Caola:2015rqy}
F.~Caola, K.~Melnikov, R.~Röntsch, and L.~Tancredi, {\it {QCD corrections to
  $W^+W^-$ production through gluon fusion}},  {\em Phys. Lett.} {\bf B754}
  (2016) 275--280, [\href{http://xxx.lanl.gov/abs/1511.0861}{{\tt
  arXiv:1511.0861}}].

\bibitem{Grazzini:2016ctr}
M.~Grazzini, S.~Kallweit, S.~Pozzorini, D.~Rathlev, and M.~Wiesemann, {\it
  {$W^+W^-$ production at the LHC: fiducial cross sections and distributions in
  NNLO QCD}},  \href{http://xxx.lanl.gov/abs/1605.0271}{{\tt arXiv:1605.0271}}.

\bibitem{Bierweiler:2012kw}
A.~Bierweiler, T.~Kasprzik, J.~H. K{\"u}hn, and S.~Uccirati, {\it {Electroweak
  corrections to W-boson pair production at the LHC}},  {\em JHEP} {\bf 11}
  (2012) 093, [\href{http://xxx.lanl.gov/abs/1208.3147}{{\tt
  arXiv:1208.3147}}].

\bibitem{Baglio:2013toa}
J.~Baglio, L.~D. Ninh, and M.~M. Weber, {\it {Massive gauge boson pair
  production at the LHC: a next-to-leading order story}},  {\em Phys. Rev.}
  {\bf D88} (2013) 113005, [\href{http://xxx.lanl.gov/abs/1307.4331}{{\tt
  arXiv:1307.4331}}].

\bibitem{Billoni:2013aba}
M.~Billoni, S.~Dittmaier, B.~Jaeger, and C.~Speckner, {\it {Next-to-leading
  order electroweak corrections to $pp \rightarrow W^+W^- \rightarrow$ 4
  leptons at the LHC in double-pole approximation}},  {\em JHEP} {\bf 12}
  (2013) 043, [\href{http://xxx.lanl.gov/abs/1310.1564}{{\tt
  arXiv:1310.1564}}].

\bibitem{Biedermann:2016guo}
B.~Biedermann, M.~Billoni, A.~Denner, S.~Dittmaier, L.~Hofer, B.~Jager, and
  L.~Salfelder, {\it {Next-to-leading-order electroweak corrections to $pp \to
  W^+W^-\to$ 4 leptons at the LHC}},
  \href{http://xxx.lanl.gov/abs/1605.0341}{{\tt arXiv:1605.0341}}.

\bibitem{Grazzini:2005vw}
M.~Grazzini, {\it {Soft-gluon effects in WW production at hadron colliders}},
  {\em JHEP} {\bf 01} (2006) 095,
  [\href{http://xxx.lanl.gov/abs/hep-ph/0510337}{{\tt hep-ph/0510337}}].

\bibitem{Wang:2013qua}
Y.~Wang, C.~S. Li, Z.~L. Liu, D.~Y. Shao, and H.~T. Li, {\it
  {Transverse-Momentum Resummation for Gauge Boson Pair Production at the
  Hadron Collider}},  {\em Phys. Rev.} {\bf D88} (2013) 114017,
  [\href{http://xxx.lanl.gov/abs/1307.7520}{{\tt arXiv:1307.7520}}].

\bibitem{Meade:2014fca}
P.~Meade, H.~Ramani, and M.~Zeng, {\it {Transverse momentum resummation effects
  in $W^+W^-$ measurements}},  {\em Phys. Rev.} {\bf D90} (2014), no.~11
  114006, [\href{http://xxx.lanl.gov/abs/1407.4481}{{\tt arXiv:1407.4481}}].

\bibitem{Grazzini:2015wpa}
M.~Grazzini, S.~Kallweit, D.~Rathlev, and M.~Wiesemann, {\it
  {Transverse-momentum resummation for vector-boson pair production at
  NNLL+NNLO}},  {\em JHEP} {\bf 08} (2015) 154,
  [\href{http://xxx.lanl.gov/abs/1507.0256}{{\tt arXiv:1507.0256}}].

\bibitem{Dawson:2013lya}
S.~Dawson, I.~M. Lewis, and M.~Zeng, {\it {Threshold resummed and approximate
  next-to-next-to-leading order results for $W^+W^-$ pair production at the
  LHC}},  {\em Phys. Rev.} {\bf D88} (2013), no.~5 054028,
  [\href{http://xxx.lanl.gov/abs/1307.3249}{{\tt arXiv:1307.3249}}].

\bibitem{Jaiswal:2014yba}
P.~Jaiswal and T.~Okui, {\it {Explanation of the $WW$ excess at the LHC by
  jet-veto resummation}},  {\em Phys. Rev.} {\bf D90} (2014), no.~7 073009,
  [\href{http://xxx.lanl.gov/abs/1407.4537}{{\tt arXiv:1407.4537}}].

\bibitem{Becher:2014aya}
T.~Becher, R.~Frederix, M.~Neubert, and L.~Rothen, {\it {Automated NNLL $+$ NLO
  resummation for jet-veto cross sections}},  {\em Eur. Phys. J.} {\bf C75}
  (2015), no.~4 154, [\href{http://xxx.lanl.gov/abs/1412.8408}{{\tt
  arXiv:1412.8408}}].

\bibitem{Monni:2014zra}
P.~F. Monni and G.~Zanderighi, {\it {On the excess in the inclusive $
  {W}^{+}{W}^{-}\ \to\ {l}^{+}{l}^{-}\nu \overline{\nu} $ cross section}},
  {\em JHEP} {\bf 05} (2015) 013,
  [\href{http://xxx.lanl.gov/abs/1410.4745}{{\tt arXiv:1410.4745}}].

\bibitem{Dawson:2016ysj}
S.~Dawson, P.~Jaiswal, Y.~Li, H.~Ramani, and M.~Zeng, {\it {Resummation of Jet
  Veto Logarithms at N$^3$LL$_a$ + NNLO for $W^+ W^-$ production at the LHC}},
  \href{http://xxx.lanl.gov/abs/1606.0103}{{\tt arXiv:1606.0103}}.

\bibitem{Frixione:2002ik}
S.~Frixione and B.~R. Webber, {\it {Matching NLO QCD computations and parton
  shower simulations}},  {\em JHEP} {\bf 0206} (2002) 029,
  [\href{http://xxx.lanl.gov/abs/hep-ph/0204244}{{\tt hep-ph/0204244}}].

\bibitem{Hamilton:2010mb}
K.~Hamilton, {\it {A positive-weight next-to-leading order simulation of weak
  boson pair production}},  {\em JHEP} {\bf 01} (2011) 009,
  [\href{http://xxx.lanl.gov/abs/1009.5391}{{\tt arXiv:1009.5391}}].

\bibitem{Hoche:2010pf}
S.~Hoche, F.~Krauss, M.~Schonherr, and F.~Siegert, {\it {Automating the POWHEG
  method in Sherpa}},  {\em JHEP} {\bf 04} (2011) 024,
  [\href{http://xxx.lanl.gov/abs/1008.5399}{{\tt arXiv:1008.5399}}].

\bibitem{Melia:2011tj}
T.~Melia, P.~Nason, R.~Rontsch, and G.~Zanderighi, {\it {W+W-, WZ and ZZ
  production in the POWHEG BOX}},  {\em JHEP} {\bf 11} (2011) 078,
  [\href{http://xxx.lanl.gov/abs/1107.5051}{{\tt arXiv:1107.5051}}].

\bibitem{Nason:2013ydw}
P.~Nason and G.~Zanderighi, {\it {$W^+ W^-$ , $W Z$ and $Z Z$ production in the
  POWHEG-BOX-V2}},  {\em Eur. Phys. J.} {\bf C74} (2014), no.~1 2702,
  [\href{http://xxx.lanl.gov/abs/1311.1365}{{\tt arXiv:1311.1365}}].

\bibitem{Bellm:2016cks}
J.~Bellm, S.~Gieseke, N.~Greiner, G.~Heinrich, S.~Platzer, C.~Reuschle, and
  J.~F. von Soden-Fraunhofen, {\it {Anomalous coupling, top-mass and
  parton-shower effects in ${W^+W^-}$ production}},
  \href{http://xxx.lanl.gov/abs/1602.0514}{{\tt arXiv:1602.0514}}.

\bibitem{Bellm:2015jjp}
J.~Bellm et~al., {\it {Herwig 7.0/Herwig++ 3.0 release note}},  {\em Eur. Phys.
  J.} {\bf C76} (2016), no.~4 196,
  [\href{http://xxx.lanl.gov/abs/1512.0117}{{\tt arXiv:1512.0117}}].

\bibitem{Cascioli:2013gfa}
F.~Cascioli, S.~H{\"o}che, F.~Krauss, P.~Maierh{\"o}fer, S.~Pozzorini, and
  F.~Siegert, {\it {Precise Higgs-background predictions: merging NLO QCD and
  squared quark-loop corrections to four-lepton + 0,1 jet production}},  {\em
  JHEP} {\bf 01} (2014) 046, [\href{http://xxx.lanl.gov/abs/1309.0500}{{\tt
  arXiv:1309.0500}}].

\bibitem{Gehrmann:2012yg}
T.~Gehrmann, S.~Hoche, F.~Krauss, M.~Schonherr, and F.~Siegert, {\it {NLO QCD
  matrix elements + parton showers in $e^+e^-$ ---> hadrons}},  {\em JHEP} {\bf
  01} (2013) 144, [\href{http://xxx.lanl.gov/abs/1207.5031}{{\tt
  arXiv:1207.5031}}].

\bibitem{Hoeche:2012yf}
S.~Hoeche, F.~Krauss, M.~Schonherr, and F.~Siegert, {\it {QCD matrix elements +
  parton showers: The NLO case}},  {\em JHEP} {\bf 04} (2013) 027,
  [\href{http://xxx.lanl.gov/abs/1207.5030}{{\tt arXiv:1207.5030}}].

\bibitem{Frederix:2011ss}
R.~Frederix, S.~Frixione, V.~Hirschi, F.~Maltoni, R.~Pittau, and P.~Torrielli,
  {\it {Four-lepton production at hadron colliders: aMC@NLO predictions with
  theoretical uncertainties}},  {\em JHEP} {\bf 02} (2012) 099,
  [\href{http://xxx.lanl.gov/abs/1110.4738}{{\tt arXiv:1110.4738}}].

\bibitem{Frederix:2012ps}
R.~Frederix and S.~Frixione, {\it {Merging meets matching in MC@NLO}},  {\em
  JHEP} {\bf 12} (2012) 061, [\href{http://xxx.lanl.gov/abs/1209.6215}{{\tt
  arXiv:1209.6215}}].

\bibitem{Alwall:2014hca}
J.~Alwall, R.~Frederix, S.~Frixione, V.~Hirschi, F.~Maltoni, O.~Mattelaer,
  H.~S. Shao, T.~Stelzer, P.~Torrielli, and M.~Zaro, {\it {The automated
  computation of tree-level and next-to-leading order differential cross
  sections, and their matching to parton shower simulations}},  {\em JHEP} {\bf
  07} (2014) 079, [\href{http://xxx.lanl.gov/abs/1405.0301}{{\tt
  arXiv:1405.0301}}].

\bibitem{Hamilton:2012rf}
K.~Hamilton, P.~Nason, C.~Oleari, and G.~Zanderighi, {\it {Merging H/W/Z + 0
  and 1 jet at NLO with no merging scale: a path to parton shower + NNLO
  matching}},  \href{http://xxx.lanl.gov/abs/1212.4504}{{\tt arXiv:1212.4504}}.

\bibitem{Kauer:2012hd}
N.~Kauer and G.~Passarino, {\it {Inadequacy of zero-width approximation for a
  light Higgs boson signal}},  {\em JHEP} {\bf 08} (2012) 116,
  [\href{http://xxx.lanl.gov/abs/1206.4803}{{\tt arXiv:1206.4803}}].

\bibitem{Nason:2004rx}
P.~Nason, {\it {A New method for combining NLO QCD with shower Monte Carlo
  algorithms}},  {\em JHEP} {\bf 0411} (2004) 040,
  [\href{http://xxx.lanl.gov/abs/hep-ph/0409146}{{\tt hep-ph/0409146}}].

\bibitem{Frixione:2007vw}
S.~Frixione, P.~Nason, and C.~Oleari, {\it {Matching NLO QCD computations with
  Parton Shower simulations: the POWHEG method}},  {\em JHEP} {\bf 0711} (2007)
  070, [\href{http://xxx.lanl.gov/abs/0709.2092}{{\tt arXiv:0709.2092}}].

\bibitem{Alioli:2010xd}
S.~Alioli, P.~Nason, C.~Oleari, and E.~Re, {\it {A general framework for
  implementing NLO calculations in shower Monte Carlo programs: the POWHEG
  BOX}},  {\em JHEP} {\bf 06} (2010) 043,
  [\href{http://xxx.lanl.gov/abs/1002.2581}{{\tt arXiv:1002.2581}}].

\bibitem{Astill:2016hpa}
W.~Astill, W.~Bizon, E.~Re, and G.~Zanderighi, {\it {NNLOPS accurate associated
  HW production}},  \href{http://xxx.lanl.gov/abs/1603.0162}{{\tt
  arXiv:1603.0162}}.

\bibitem{Frederix:2015fyz}
R.~Frederix and K.~Hamilton, {\it {Extending the MINLO method}},  {\em JHEP}
  {\bf 05} (2016) 042, [\href{http://xxx.lanl.gov/abs/1512.0266}{{\tt
  arXiv:1512.0266}}].

\bibitem{Hamilton:2013fea}
K.~Hamilton, P.~Nason, E.~Re, and G.~Zanderighi, {\it {NNLOPS simulation of
  Higgs boson production}},  {\em JHEP} {\bf 1310} (2013) 222,
  [\href{http://xxx.lanl.gov/abs/1309.0017}{{\tt arXiv:1309.0017}}].

\bibitem{Hamilton:2015nsa}
K.~Hamilton, P.~Nason, and G.~Zanderighi, {\it {Finite quark-mass effects in
  the NNLOPS POWHEG+MiNLO Higgs generator}},  {\em JHEP} {\bf 05} (2015) 140,
  [\href{http://xxx.lanl.gov/abs/1501.0463}{{\tt arXiv:1501.0463}}].

\bibitem{Karlberg:2014qua}
A.~Karlberg, E.~Re, and G.~Zanderighi, {\it {NNLOPS accurate Drell-Yan
  production}},  {\em JHEP} {\bf 1409} (2014) 134,
  [\href{http://xxx.lanl.gov/abs/1407.2940}{{\tt arXiv:1407.2940}}].

\bibitem{Alioli:2012fc}
S.~Alioli, C.~W. Bauer, C.~J. Berggren, A.~Hornig, F.~J. Tackmann, et~al., {\it
  {Combining Higher-Order Resummation with Multiple NLO Calculations and Parton
  Showers in GENEVA}},  \href{http://xxx.lanl.gov/abs/1211.7049}{{\tt
  arXiv:1211.7049}}.

\bibitem{Alioli:2013hqa}
S.~Alioli, C.~W. Bauer, C.~Berggren, F.~J. Tackmann, J.~R. Walsh, and
  S.~Zuberi, {\it {Matching Fully Differential NNLO Calculations and Parton
  Showers}},  {\em JHEP} {\bf 06} (2014) 089,
  [\href{http://xxx.lanl.gov/abs/1311.0286}{{\tt arXiv:1311.0286}}].

\bibitem{Alioli:2015toa}
S.~Alioli, C.~W. Bauer, C.~Berggren, F.~J. Tackmann, and J.~R. Walsh, {\it
  {Drell-Yan Production at NNLL'+NNLO Matched to Parton Showers}},
  \href{http://xxx.lanl.gov/abs/1508.0147}{{\tt arXiv:1508.0147}}.

\bibitem{Alioli:2016wqt}
S.~Alioli, C.~W. Bauer, S.~Guns, and F.~J. Tackmann, {\it {Underlying event
  sensitive observables in Drell-Yan production using GENEVA}},
  \href{http://xxx.lanl.gov/abs/1605.0719}{{\tt arXiv:1605.0719}}.

\bibitem{Hoeche:2014aia}
S.~H{\"o}che, Y.~Li, and S.~Prestel, {\it {Drell-Yan lepton pair production at
  NNLO QCD with parton showers}},  {\em Phys. Rev.} {\bf D91} (2015), no.~7
  074015, [\href{http://xxx.lanl.gov/abs/1405.3607}{{\tt arXiv:1405.3607}}].

\bibitem{Hoche:2014dla}
S.~H{\"o}che, Y.~Li, and S.~Prestel, {\it {Higgs-boson production through gluon
  fusion at NNLO QCD with parton showers}},  {\em Phys. Rev.} {\bf D90} (2014),
  no.~5 054011, [\href{http://xxx.lanl.gov/abs/1407.3773}{{\tt
  arXiv:1407.3773}}].

\bibitem{Hamilton:2012np}
K.~Hamilton, P.~Nason, and G.~Zanderighi, {\it {MINLO: Multi-Scale Improved
  NLO}},  {\em JHEP} {\bf 1210} (2012) 155,
  [\href{http://xxx.lanl.gov/abs/1206.3572}{{\tt arXiv:1206.3572}}].

\bibitem{Alwall:2007st}
J.~Alwall, P.~Demin, S.~de~Visscher, R.~Frederix, M.~Herquet, F.~Maltoni,
  T.~Plehn, D.~L. Rainwater, and T.~Stelzer, {\it {MadGraph/MadEvent v4: The
  New Web Generation}},  {\em JHEP} {\bf 09} (2007) 028,
  [\href{http://xxx.lanl.gov/abs/0706.2334}{{\tt arXiv:0706.2334}}].

\bibitem{Campbell:2012am}
J.~M. Campbell, R.~K. Ellis, R.~Frederix, P.~Nason, C.~Oleari, and C.~Williams,
  {\it {NLO Higgs Boson Production Plus One and Two Jets Using the POWHEG BOX,
  MadGraph4 and MCFM}},  {\em JHEP} {\bf 07} (2012) 092,
  [\href{http://xxx.lanl.gov/abs/1202.5475}{{\tt arXiv:1202.5475}}].

\bibitem{Cullen:2014yla}
G.~Cullen et~al., {\it {G$\scriptsize{O}$S$\scriptsize{AM}$-2.0: a tool for
  automated one-loop calculations within the Standard Model and beyond}},  {\em
  Eur. Phys. J.} {\bf C74} (2014), no.~8 3001,
  [\href{http://xxx.lanl.gov/abs/1404.7096}{{\tt arXiv:1404.7096}}].

\bibitem{Melia:2010bm}
T.~Melia, K.~Melnikov, R.~Rontsch, and G.~Zanderighi, {\it {Next-to-leading
  order QCD predictions for $W^+W^+jj$ production at the LHC}},  {\em JHEP}
  {\bf 12} (2010) 053, [\href{http://xxx.lanl.gov/abs/1007.5313}{{\tt
  arXiv:1007.5313}}].

\bibitem{Ball:2014uwa}
{\bf NNPDF} Collaboration, R.~D. Ball et~al., {\it {Parton distributions for
  the LHC Run II}},  {\em JHEP} {\bf 04} (2015) 040,
  [\href{http://xxx.lanl.gov/abs/1410.8849}{{\tt arXiv:1410.8849}}].

\bibitem{Sjostrand:2006za}
T.~Sj{\"o}strand, S.~Mrenna, and P.~Z. Skands, {\it {PYTHIA 6.4 Physics and
  Manual}},  {\em JHEP} {\bf 05} (2006) 026,
  [\href{http://xxx.lanl.gov/abs/hep-ph/0603175}{{\tt hep-ph/0603175}}].

\bibitem{Sjostrand:2007gs}
T.~Sj{\"o}strand, S.~Mrenna, and P.~Z. Skands, {\it {A Brief Introduction to
  PYTHIA 8.1}},  {\em Comput. Phys. Commun.} {\bf 178} (2008) 852--867,
  [\href{http://xxx.lanl.gov/abs/0710.3820}{{\tt arXiv:0710.3820}}].

\bibitem{Sjostrand:2014zea}
T.~Sj{\"o}strand, S.~Ask, J.~R. Christiansen, R.~Corke, N.~Desai, P.~Ilten,
  S.~Mrenna, S.~Prestel, C.~O. Rasmussen, and P.~Z. Skands, {\it {An
  Introduction to PYTHIA 8.2}},  {\em Comput. Phys. Commun.} {\bf 191} (2015)
  159--177, [\href{http://xxx.lanl.gov/abs/1410.3012}{{\tt arXiv:1410.3012}}].

\bibitem{Luisoni:2013kna}
G.~Luisoni, P.~Nason, C.~Oleari, and F.~Tramontano, {\it {$HW^{\pm}$/HZ + 0 and
  1 jet at NLO with the POWHEG BOX interfaced to GoSam and their merging within
  MiNLO}},  {\em JHEP} {\bf 10} (2013) 083,
  [\href{http://xxx.lanl.gov/abs/1306.2542}{{\tt arXiv:1306.2542}}].

\bibitem{Cacciari:2008gp}
M.~Cacciari, G.~P. Salam, and G.~Soyez, {\it {The
  Anti-$k_{\mathrm{\scriptscriptstyle{T}}}$ jet clustering algorithm}},  {\em
  JHEP} {\bf 04} (2008) 063, [\href{http://xxx.lanl.gov/abs/0802.1189}{{\tt
  arXiv:0802.1189}}].

\bibitem{Cacciari:2011ma}
M.~Cacciari, G.~P. Salam, and G.~Soyez, {\it {FastJet User Manual}},  {\em
  Eur.Phys.J.} {\bf C72} (2012) 1896,
  [\href{http://xxx.lanl.gov/abs/1111.6097}{{\tt arXiv:1111.6097}}].

\end{thebibliography}\endgroup

\end{document}